\documentclass{scienceadvhack}

\usepackage{times}
\usepackage{graphicx}
\usepackage{amsmath}
\usepackage{amssymb}
\usepackage{url}
\usepackage{hyperref}

\usepackage{newfloat}
\usepackage{caption}
\usepackage{float}

\DeclareFloatingEnvironment[
    fileext=lof,
    name=Supplementary Figure,
    placement=tbhp,
    within=none
]{suppfigure}

\DeclareFloatingEnvironment[
    fileext=lot,
    name=Supplementary Table,
    placement=tbhp,
    within=none
]{supptable}

\newcommand{\aj}{Astron. J.}   
\newcommand{\apj}{Astrophys. J.}   
\newcommand{\apjl}{Astrophys. J. Lett.}   
\newcommand{\apjs}{Astrophys. J. Suppl. Ser.}   
\newcommand{\aap}{Astron. Astrophys.}   
\newcommand{\mnras}{Mon. Not. R. Astron. Soc.}   
\newcommand{\nat}{Nature} 
\newcommand{\pasp}{Publ. Astron. Soc. Pac.}   
\newcommand{\ssr}{Space Sci. Rev.}   

\newcounter{lastnote}

\title{A persistent bow shock in a diskless magnetised accreting white dwarf}

\author
{Krystian I{\l}kiewicz$^{1,2\dagger}$,
Simone Scaringi$^{2,3\ast\dagger}$, 
Domitilla de Martino$^{3}$,
Christian Knigge$^{4}$, 
Sara E. Motta$^{5,6}$, 
Nanda Rea$^{7,8}$, 
David Buckley$^{9,10,11}$, 
Noel Castro Segura$^{12}$, 
Paul J. Groot$^{13,9,10}$, 
Anna F. McLeod$^{2,14}$, 
Luke T. Parker$^{6}$, 
Martina Veresvarska$^{2,7,8}$. 
\\
\\
\normalsize{$^{1}$Nicolaus Copernicus Astronomical Center, Polish Academy of Sciences, ul. Bartycka 18, 00-716 Warsaw, Poland}\\
\normalsize{$^{2}$Centre for Extragalactic Astronomy, Department of Physics, Durham University, DH1 3LE, United Kingdom}\\
\normalsize{$^{3}$INAF-Osservatorio Astronomico di Capodimonte, Salita Moiariello 16, I-80131 Naples, Italy}\\
\normalsize{$^{4}$School of Physics and Astronomy, University of Southampton, Highfield, Southampton SO17 1BJ, United Kingdom}\\
\normalsize{$^{5}$INAF-Osservatorio Astronomico di Brera, via E. Bianchi 46, I-23807, Merate, Italy}\\
\normalsize{$^{6}$University of Oxford, Department of Physics, Astrophysics, Denys Wilkinson Building, Keble Road, OX1 3RH, Oxford, United Kingdom}\\
\normalsize{$^{7}$Institute of Space Sciences (ICE, CSIC), Campus UAB, Carrer de Can Magrans s/n, 08193, Barcellona, Spain}\\
\normalsize{$^{8}$Institut d’Estudis Espacials de Catalunya (IEEC), Esteve Terradas 1, RDIT Building, 08860, Castelldefels, Spain}\\
\normalsize{$^{9}$South African Astronomical Observatory,  PO Box 9, Observatory, 7935, Cape Town, South Africa}\\
\normalsize{$^{10}$Department of Astronomy, University of Cape Town, Private Bag X3, Rondebosch, 7701, South Africa}\\
\normalsize{$^{11}$Department of Physics, University of the Free State, PO Box 339, 9300, Bloemfontein, South Africa}\\
\normalsize{$^{12}$Department of Physics, University of Warwick, Gibbet Hill Road, Coventry, CV4 7AL, United Kingdom}\\
\normalsize{$^{13}$Department of Astrophysics/IMAPP, Radboud University, P.O. 9010, 6500 GL, Nĳmegen, The Netherlands}\\
\normalsize{$^{14}$Institute for Computational Cosmology, Department of Physics, Durham University, South Road, DH1 3LE, United Kingdom}\\
\\
\normalsize{$^\ast$Corresponding author. E-mail: simone.scaringi@durham.ac.uk}
\\
\normalsize{$^\dagger$These authors contributed equally to this work.}
\\
\normalsize{Submitted on 28 May 2025. Accepted for publication in \textit{Nature Astronomy} on 20 November 2025.}
}

\date{}

\begin{document}

\maketitle 

\begin{abstract}
Stellar bow shocks are formed when an outflow interacts with the interstellar medium. In white dwarfs accreting from a binary companion, outflows are associated with either strong winds from the donor star, the accretion disk, or a thermonuclear runaway explosion on the white dwarf surface. To date, only six accreting white dwarfs are known to harbour disk-wind driven bow shocks that are not associated to thermonuclear explosions. Here, we report the discovery of a bow shock associated with a high-proper-motion disk-less accreting white dwarf, 1RXS~J052832.5+283824. We show that the white dwarf has a strong magnetic field in the range B$\approx$42--45~MG, making RXJ0528+2838 the a bonafide known polar-type cataclysmic variable harbouring a bow shock. The resolved bow shock is shown to be inconsistent with a past thermonuclear explosion, or being inflated by a donor wind, ruling out all accepted scenarios for inflating a bow shock around this system. Modelling of the energetics reveals that the observed bow shock requires a persistent power source with a luminosity significantly exceeding the system accretion energy output. This implies the presence of a powerful, previously unrecognized energy loss mechanism — potentially tied to magnetic activity — that may operate over sufficiently long timescales to influence the course of binary evolution.
\end{abstract}

1RXS~J052832.5+283824 (hereafter RXJ0528+2838) is a short-period polar-type Cataclysmic Variable (CV) with an orbital period of 80 minutes, which puts it at the theoretical orbital period minimum for such systems\cite{2015AstBu..70..460G}. The \textit{Gaia} parallax measurement indicates a system distance of $224.2 \pm 4.1$~pc$^[$\cite{2021AJ....161..147B}$^]$. Spectropolarimetric data and emission-line diagnostics confirm its polar classification\cite{2016AstBu..71...95B,2021ApJS..257...65S}. We confirm\cite{2016AstBu..71...95B} the magnetic field strength in the range \( B \approx 42\!-\!45~\mathrm{MG} \) (see \textit{Methods}) from the cyclotron harmonic structures in the Multi Unit Spectroscopic Explorer (MUSE) spectrum (see \textit{Methods}), consistent with a strongly magnetized accreting white dwarf.

Recent \textit{XMM-Newton} observations also confirm\cite{2016AstBu..71...95B} RXJ0528+2838 to display all the classic properties of polars (see \textit{Methods}). We identify a previously unknown nebular structure surrounding this magnetized accreting white dwarf using H$\alpha$ imaging data from the INT Galactic Plane Survey (IGAPS)\cite{2005MNRAS.362..753D,2021A&A...655A..49G}. The nebula exhibits a distinct bow shock morphology with a faint trailing structure. Follow-up observations with MUSE (see \textit{Methods}) confirm the presence of this bow shock, revealing a complex and stratified emission-line structure. 

\begin{figure*}
   \centering
   \includegraphics[width=0.99\textwidth]{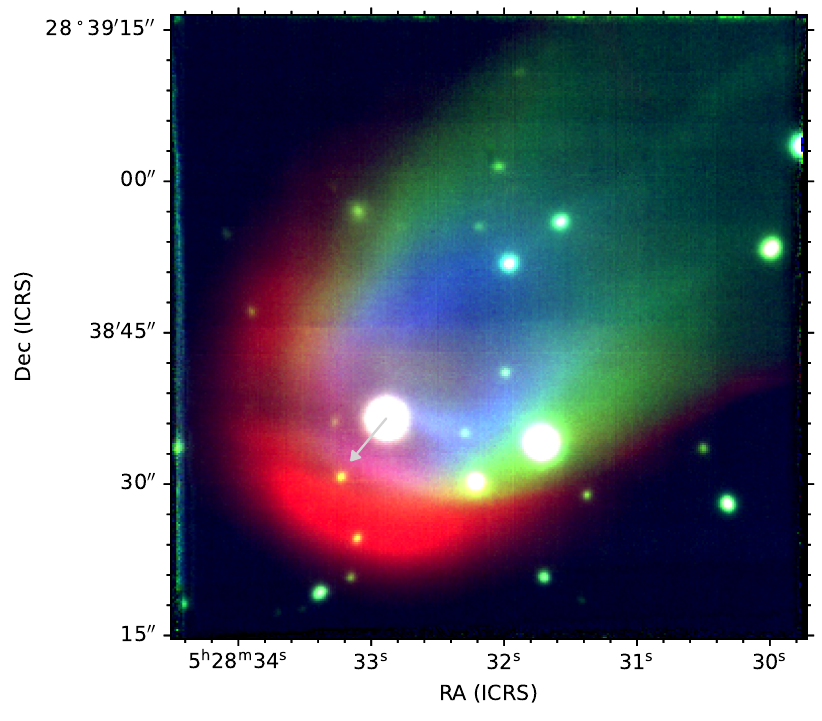}
      \caption{\textbf{False-colour image of RXJ0528+2838 and its surrounding nebula.} The red, green, and blue channels correspond to the H$\alpha$, {
      \sc [N~II]}~6548\AA, and {\sc [O~III]}~5007\AA\, lines, respectively, extracted using a top-hat filter from the MUSE datacube. The gray arrow indicates the proper motion of RXJ0528+2838\cite{2021A&A...649A...1G}. North is up, East is left.}
         \label{fig:rgb}
   \end{figure*}

\begin{figure*}
\centering
   \includegraphics[width=0.99\columnwidth]{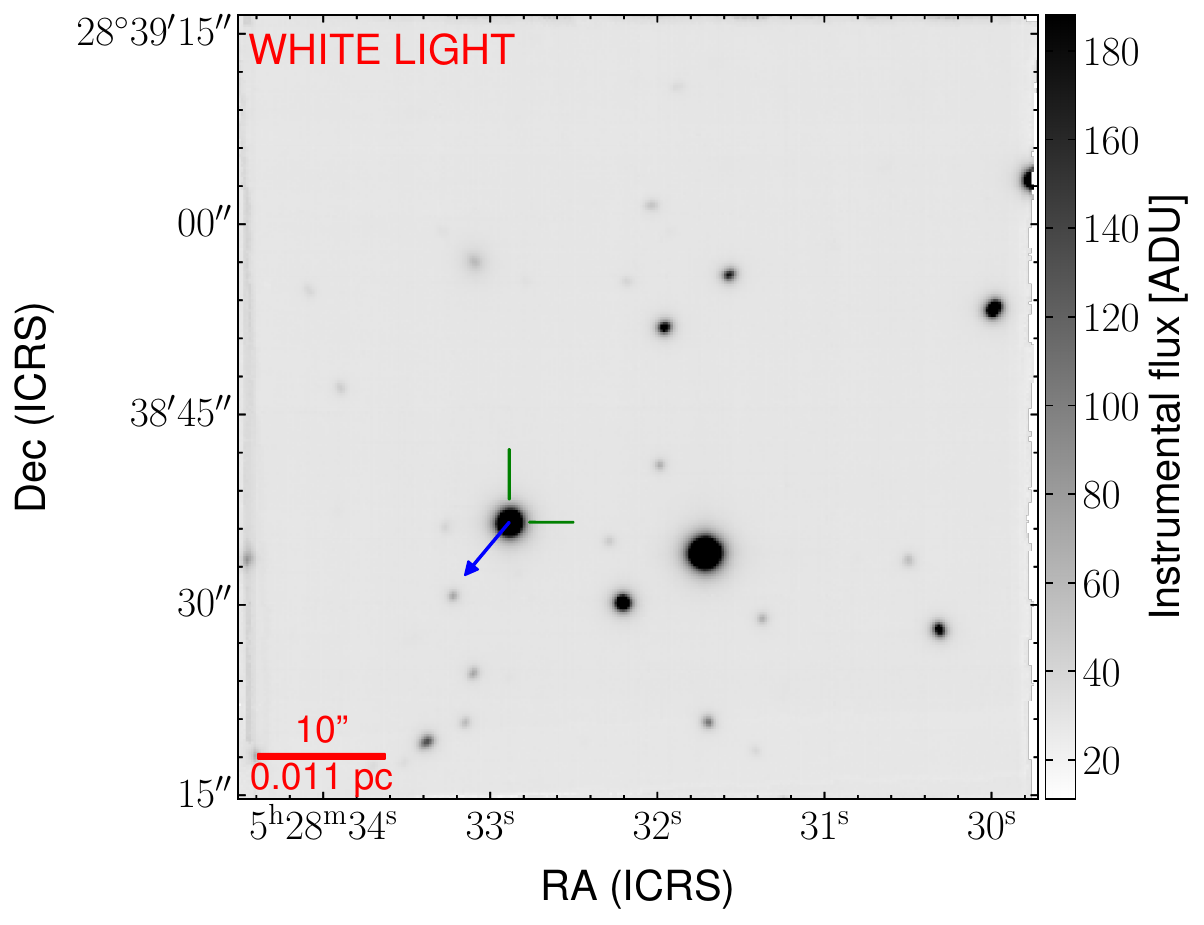}
   \includegraphics[width=0.99\columnwidth]{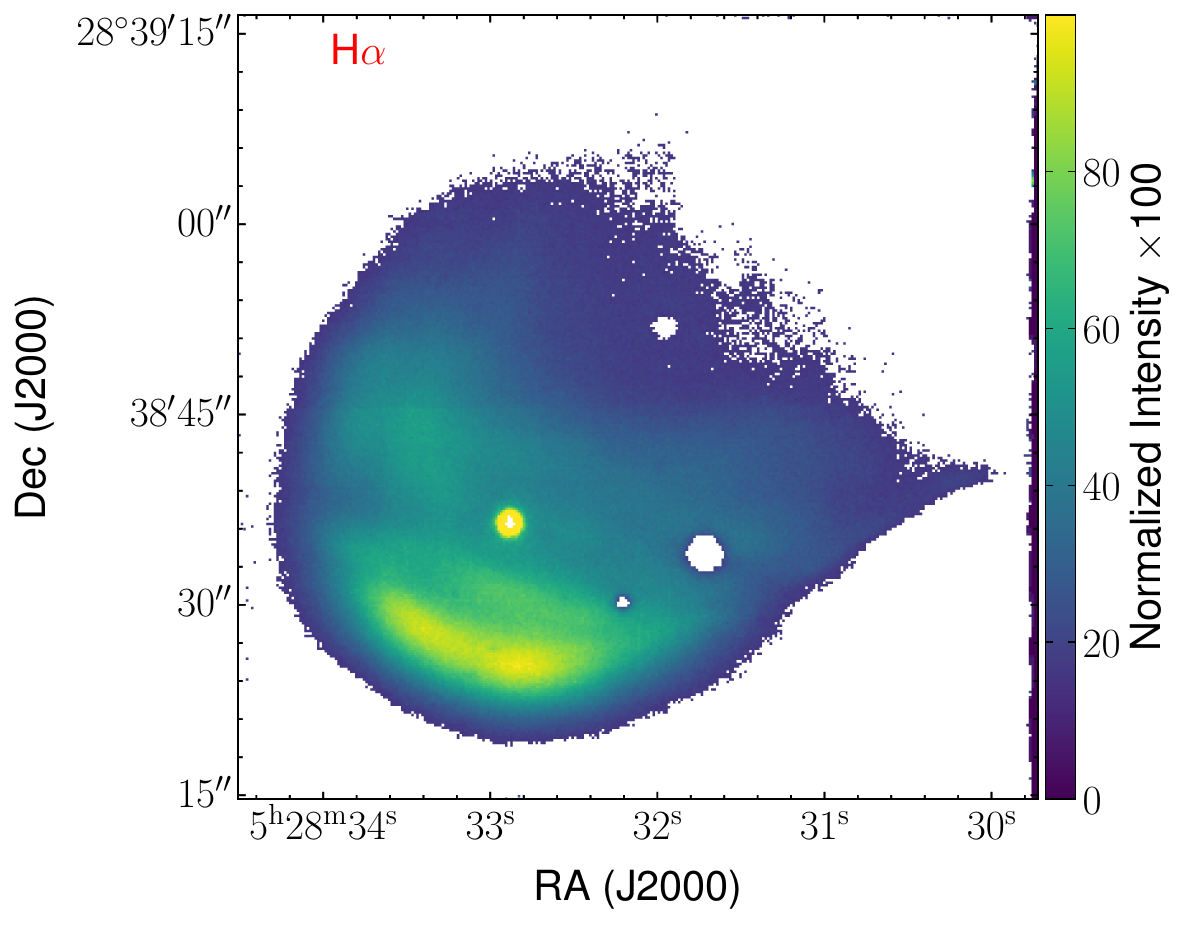}
   \includegraphics[width=0.99\columnwidth]{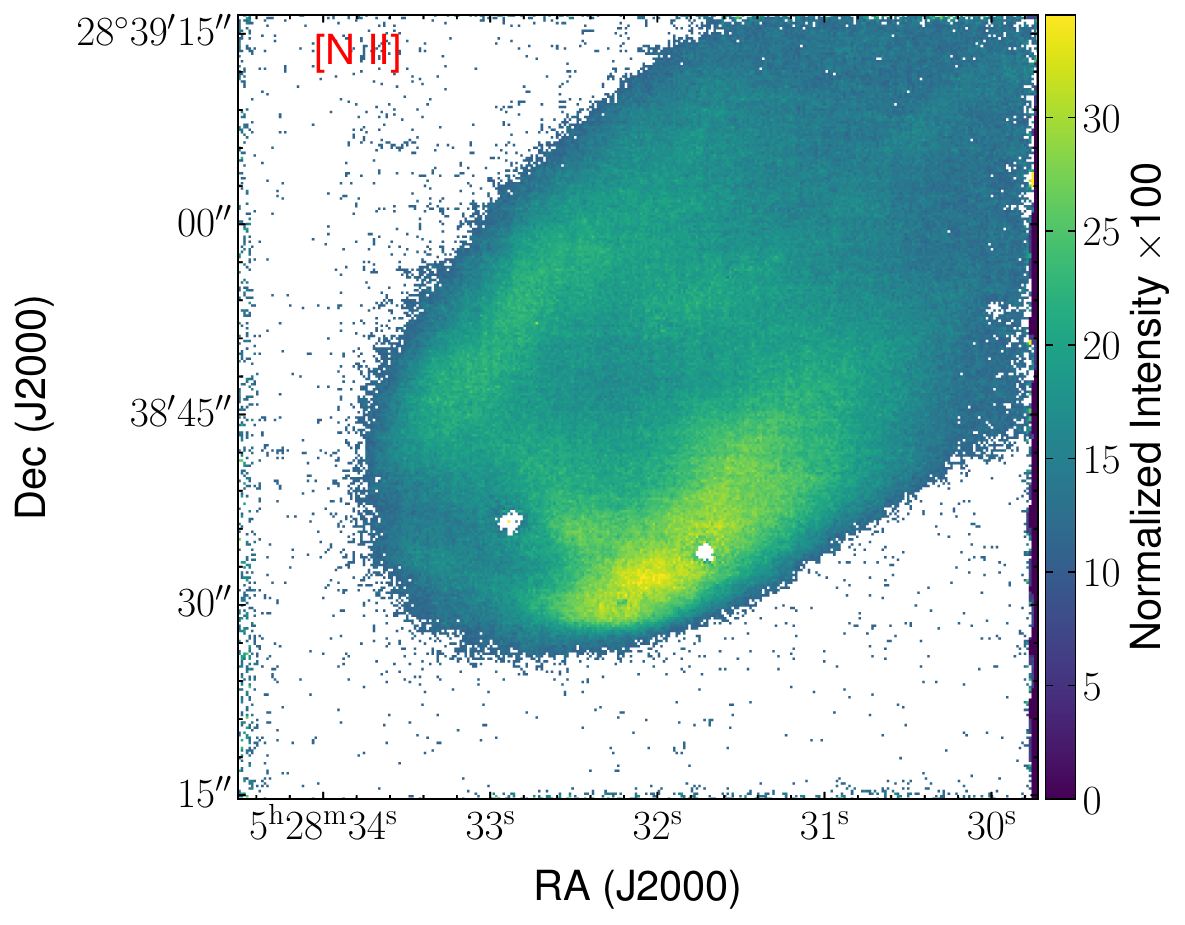}
   \includegraphics[width=0.99\columnwidth]{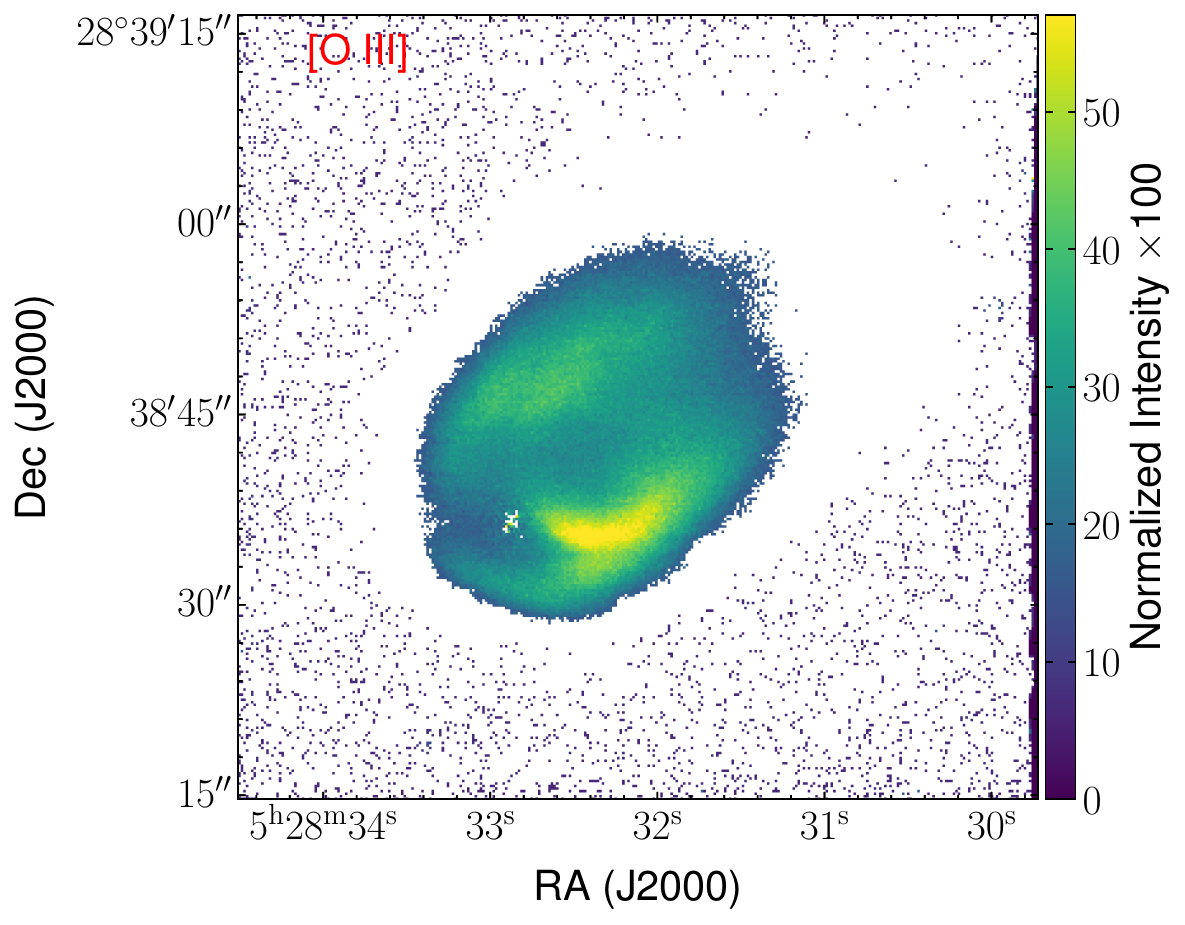}
   \includegraphics[width=0.99\columnwidth]{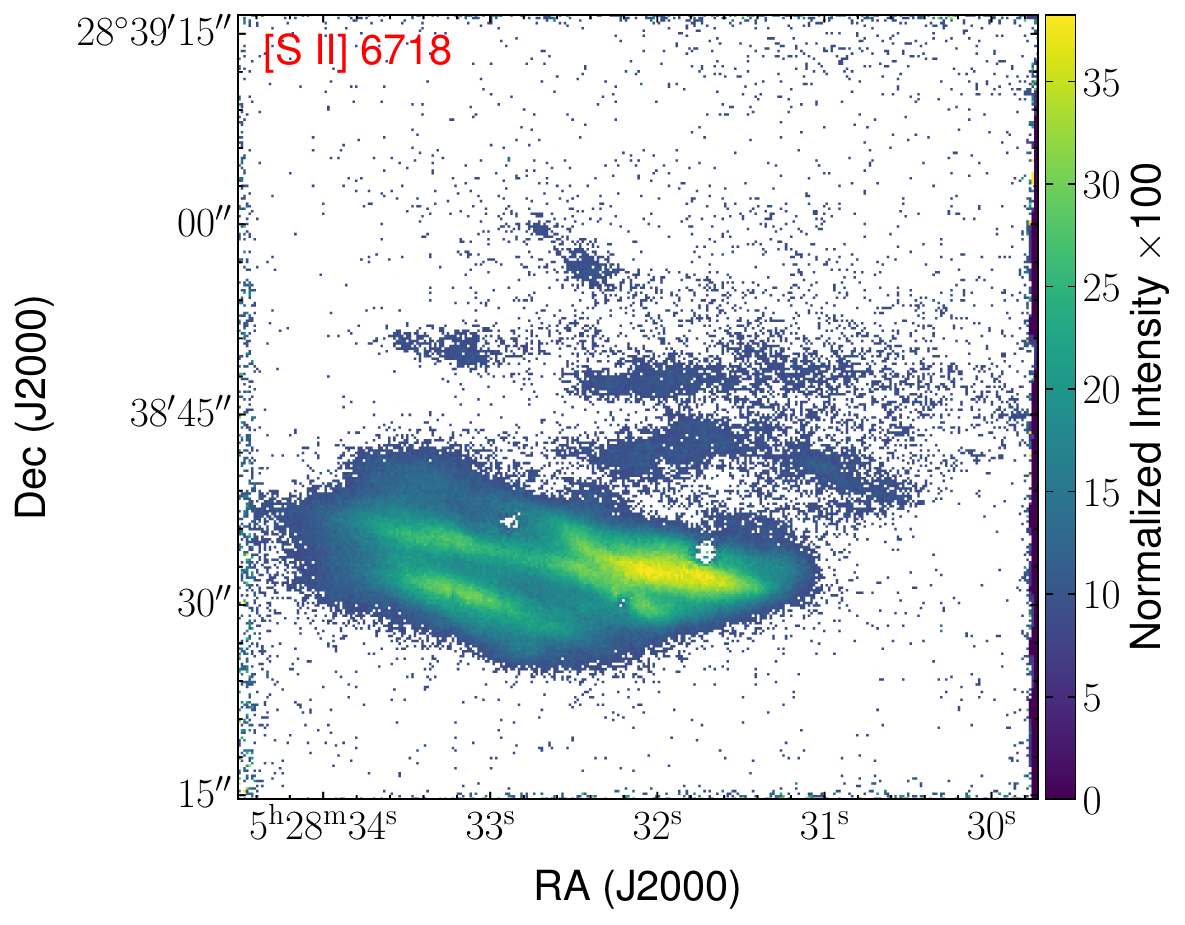}
   \includegraphics[width=0.99\columnwidth]{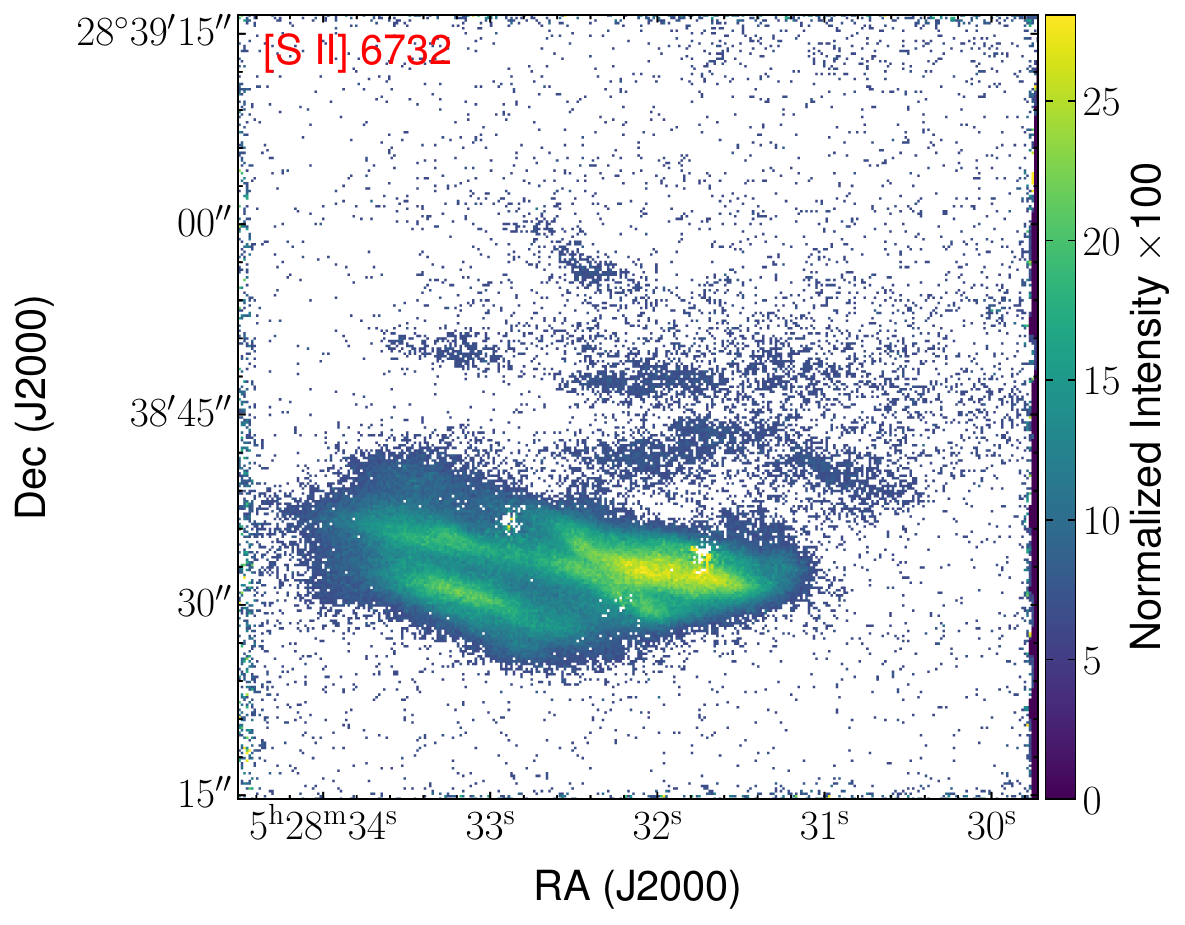}
      \caption{\textbf{MUSE observation of RXJ0528+2838 and its extended emission.} Top left: white light image of the field with the CV marked with green lines and the direction of its proper motion with a blue arrow. Intensities of measured emission lines are displayed in the other panels. All colour scales are normalised to the maximum $H\alpha$ intensity in the MUSE datacube after being multiplied by a factor of 100 for display purposes.}
\label{fig:lines}
\end{figure*}

The morphology of the bow shock aligns with the proper motion of RXJ0528+2838 (Fig.~\ref{fig:rgb} and Fig.~\ref{fig:lines}), establishing a direct physical association between the object and the nebula. The shape of the bow shock varies greatly between the observed emission lines, implying a strong stratification of the nebula. In particular, the size of the bow shock differs depending on the observed emission line. Specifically, the apex of the shock lies at approximately 16.7~arcsec (3800~AU, projected) from the CV in the hydrogen lines of the Balmer series transitions, 10.5~arcsec (2400~AU, projected) in {\sc [N~II]}, and 6.25~arcsec (1400~AU, projected) in {\sc [O~III]}. In contrast, {\sc [S~II]} emission lacks a clear bow shock geometry, with diffuse emission filling the space between the star and the H$\alpha$ apex (see \textit{Methods}). Despite the system's proper motion being nearly exactly South East, the emission across all lines is systematically brighter to the South West, suggesting anisotropic outflow forming the bow shock. In particular, {\sc [O~III]} appears to trace a directed, non-spherical outflow originating from the CV.

The presence of a bow shock nebula around RXJ0528+2838 is unexpected. Among the population of accreting white dwarfs, approximately 36\% host strongly magnetized white dwarfs\cite{2020MNRAS.494.3799P,2025PASP..137a4201R}, yet extended nebular structures associated with polars remain virtually non-existent, with the exception of a nova shell related to a past thermonuclear explosion around the asynchronous polar V1500~Cyg\cite{1995MNRAS.276..353S}. All other bow shocks know to date associated to accreting white dwarf systems are thought to be powered by disk winds (e.g.\cite{1987A&A...181..373K,2018PASP..130i4201B,2019MNRAS.486.2631H,2021MNRAS.501.1951C,2024AJ....168..249B}), while those around accreting neutron stars are thought to be powered by pulsar wind nebulae (e.g.\cite{brownsberger14,bhalerao19}). The lack of an accretion disk in RXJ0528+2838 rules out disk winds as the power source inflating the bow shock, while the white dwarf compact accretor and associated magnetic field strength rules out a pulsar wind as the energy source, leaving the observed bow shock around RXJ0528+2838 unexplained.

The RXJ0528+2838 nebula morphology and kinematics distinguish it from classical nova remnants. A nova, at the source distance of 224\,pc would have reached $\sim$3.5~mag at maximum, making it visible to the naked eye and likely to be discovered, unless it occurred during a time of year when it was positioned too close to the Sun in the sky. Nonetheless, nova shells expand independently of their progenitor systems and generally retain spherical or elliptical symmetry, even when strong interaction with the interstellar matter is present\cite{2016A&A...595A..64H,2024ApJ...972L..14I,2024A&A...681A.106C}, although this is not always the case\cite{celedon25}. In contrast, the nebula around RXJ0528+2838 is sharply asymmetric, shaped into a bow shock aligned with the CV proper motion, and exhibits a smooth {\sc [N~II]}~6548\AA\, emission tail trailing behind the system. This tail implies a continuous or quasi-continuous outflow over the past $\sim$1000 years or more (see \textit{Methods}), inconsistent with the episodic ejection of material expected from classical nova eruptions. Moreover, nova shells are known to show a radial velocity structure indicating quasi-spherical expansion\cite{2007MNRAS.380..175V,2022MNRAS.517.2567S,2024A&A...681A.106C}, while in MUSE data of RXJ0528+2838 the radial velocities remain constant across the nebula (see \textit{Methods}). The bow shock is thus a long-lived, possibly quasi-steady, structure produced by ongoing energy injection (potentially via mass loss) from RXJ0528+2838 - implying an as-yet uncharacterised, non-episodic mode of outflow in synchronized magnetic CVs.

The energy required to sustain the bow shock, $L_{b}$, can be estimated using the analytical equation developed for wind–interstellar medium interactions\cite{1977ApJ...218..377W,2021MNRAS.501.1951C}:
\begin{equation}
    L_{b} = 14.85 \pi (R_b \sec\Theta)^2 \rho_{ISM} v_{ISM}^3,
\end{equation}

\noindent where $R_b$ is the size of the bow shock projected onto the plane of the sky, $\Theta$ is the inclination of the bow shock relative to the plane of sky, $\rho_{ISM}$ is the density of the interstellar medium, and $v_{ISM}$ is the velocity of the star relative to the interstellar medium. We adopt a projected size of 16.7~arcsec and a RXJ0528+2838 distance of 224~pc$^[$\cite{2021AJ....161..147B}$^]$, yielding $R_b = 0.018$~pc. Using the proper motion, radial velocity, and distance of RXJ0528+2838 (see \textit{Methods}), we derive an inclination of $\Theta = 64^{\circ}$ and $v_{ISM} = 142$~km s$^{-1}$.  We estimated the electron density of the interstellar medium at the position and distance of RXJ0528+2838 as $n_{e,ISM} = 0.017327$~cm$^{-3}$ using a model for the distribution of free electrons in the Galaxy\cite{2017ApJ...835...29Y}.
This electron density is consistent with a warm-neutral medium, which has an ionization fraction of $\sim 8\%$\cite{2013ApJ...773L..11F,2013ApJ...764...25J}. Combining the ionization fraction with the electron density, we estimate the mass density of the interstellar medium around RXJ0528+2838 as $\rho_{ISM} = 3.6 \times 10^{-25}$~g cm$^{-3}$. This yields a required power input of $L_{b} \approx  8.2 \times 10^{32}$~erg s$^{-1}$.

Bow shocks not related to nova outbursts have been identified in only six CVs\cite{1987A&A...181..373K,2018PASP..130i4201B,2019MNRAS.486.2631H,2024AJ....168..249B,2025AJ....170...78B}. These structures are generally attributed to either accretion disk winds or enhanced stellar winds from the donor star\cite{2019MNRAS.486.2631H,2021MNRAS.501.1951C}. However, since RXJ0528+2838 is a polar it lacks an accretion disk\cite{1990SSRv...54..195C}, ruling out disk winds as a viable origin for the observed mass loss and energy injection.  
If the bow shock were to be powered by an outflow — such as a wind from the donor star - it would supply a kinetic power:
\begin{equation}\label{eq:kinetic}
    L_{o} = \frac{1}{2} \dot{M}_o v_{o}^2,
\end{equation}
where $\dot{M}_o$ is the mass-loss rate through the outflow and $v_{o}$ is the outflow velocity. For a typical red-dwarf wind mass-loss rate of $\sim 10^{-14}$~M$_\odot$yr$^{-1}$, the required wind velocity would be of the order of $10^5$~km s$^{-1}$, also excluding the mass donor wind in RXJ0528+2838 as the origin of the bow shock.

Another possible energy source that could power the bow shock might be electromagnetic radiation provided by the WD spin-down. Currently the WD spin is locked with the orbit (see \textit{Methods}) and we can recover an upper limit on its spin-down.
Assuming a white dwarf mass $M_{\text{WD}} = 0.8$~M$_\odot$$^[$\cite{2022MNRAS.510.6110P}$^]$ and radius $R_{\text{WD}} = 0.01$~R$_\odot$, and assuming synchronization $\dot{P}_{\text{spin}} = \dot{P}_{\text{orb}}$, while adopting an upper limit on the spin period change of $\dot{P}_{\text{spin}} < 2 \times 10^{-9}$~s~s$^{-1}$ (see \textit{Methods}), we constrain the spin-down luminosity to $L_{\text{spin}} < 2 \times 10^{32}$~erg s$^{-1}$. This assumes an unrealistic 100\% efficiency and is yet a factor of at least 4 below the energy required to sustain the observed bow shock. As RXJ0528+2838 appears to be fully synchronized we can consider the energy stored in the binary instead. For an assumed donor mass of $M_{\text{donor}} = 0.1$~M$_\odot$, the required $\dot{P}_{\text{orb}}$ would have to be $4\times10^{-12}$~s~s$^{-1}$. Although this is well below our detection threshold, we note this is about one order of magnitude larger than that predicted from gravitational wave radiation, and that there is currently no known theoretical framework for how binary inspiral energy loss could be transferred to power the observed bow shock.

Considering the total magnetic energy of the WD of $U_{\text{mag}}\approx1.6\times10^{41}$\,erg and assuming an unreasonably fast decay (for a WD) of 1\,Gyr, the output cannot provide a power larger than $\sim3.4\times10^{24}$\,erg\,s$^{-1}$, still insufficient, also implying that the luminosity cannot be powered by magnetic energy alone. We also considered the possibility of the system containing a pulsar, which could power such a bow shock via the interaction of its relativistic wind with the ISM. We performed MeerKAT observations of the field to detect a possible radio pulsar hosted in a triple system. The derived upper limits of 38$\mu$Jy at the distance of RXJ0528+2838 yields a radio power of $2.9\times10^{24}$\,erg\,s$^{-1}$ (see \textit{Methods}), excluding any radio pulsar at the source distance unless the radio emission is anisotropic and pointed elsewhere relative to our line of sight. A powerful pulsar in the line-of-sight at a much further distance can be excluded by the strong correlation of the bow shock with the CV proper motion.

Regardless of the driving mechanism the formation and ongoing presence of the bow shock requires a persistent energy source with an average luminosity of $L_{b} \approx  8.2 \times 10^{32}$~erg s$^{-1}$. A CV with an orbital period similar to that of RXJ0528+2838 is expected to exhibit a mass-transfer rate of $5 \times 10^{-11}$~M$_\odot$ yr$^{-1}$$^[$\cite{2011ApJS..194...28K}$^]$, corresponding to an accretion luminosity of $L_{acc} = 2.4 \times 10^{32}$~erg s$^{-1}$. The energy required to energize the bow shock thus also exceeds the system accretion luminosity by a factor of $\sim$3. Combined with the inferred minimum age of the bow shock ($>1000$~years, see \textit{Methods}), this energy excess indicates a long-lived, energetically significant, outflow that cannot be explained by standard mass-transfer or mass-outflow models. 

Given that all established bow shock formation scenarios fail to explain the observations of RXJ0528+2838, we can only speculate on a novel energy injection mechanism into the ISM from this system. Rather than tapping into gravitational potential energy we can consider the power source responsible for inflating the bow shock to be tapping into the strong magnetic stored energy density of the WD and ask how long this might be depleted at the power required to inflate the bow shock. As the observed power requirement of $P=L_{bow}\approx 8.2 \times 10^{32}$~erg s$^{-1}$ the magnetic energy $U_{mag}\approx 1.6 \times 10^{41}$erg would be depleted within $\approx600$ years, in tension with the observed bow shock tail requiring a sustained energy source for at least 1000 years. One possible solution to this may be that the WD in RXJ0528+2838 possessed a much higher magnetic field of $\approx10^{9}$G in the past. This larger energy reservoir of $U_{mag}\approx 10^{43}$erg could then sustain a power output comparable to $L_{bow}$ for a few thousand years at the expense of decreasing the WD magnetic field strength. However, aside from being entirely speculative, this scenario implies not only that we are observing RXJ0528+2838 during a rare and short evolutionary phase, but that this phase has ended very recently, possibly arguing against this hypothesis. Furthermore, it is not obvious how the WD magnetic energy would be extracted in this scenario. One possibility could be quasi-continuous reconfiguration of the magnetic field, including the possibility of the WD multipole reconfiguration\cite{mason24}. Alternatively the WD in RXJ0528+2838 may have been asynchronously rotating with respect to the orbit in the not so distant past, extracting magnetic energy from the WD through binary interaction torques. If this were the case, RXJ0528+2838 would have synchronised very recently and would be a direct progenitor of so-called asynchronous polars, of which less than 10 have been discovered to date. Extended emission has been searched for at optical wavelengths in six confirmed asynchronous polars without any detection to date\cite{pagnotta16}. This is also an argument against this speculative hypothesis, although we note that all known asynchronous polars are at substantially larger distances than RXJ0528+2838.

From an observational point of view we can comment on other somewhat similar systems that seemingly exhibit energy losses that remain to be fully explained. These include, for example, the prototypical polar AM Her with a reported radio luminosity of $1.8\times10^{25}$\,erg\,s$^{-1}$$^[$\cite{ridder23}$^]$, the non-accreting white dwarf pulsar analogue AR Sco with a radio luminosity of $\approx10^{27}$\,erg\,s$^{-1}$$^[$\cite{marsh16}$^]$, as well as the non-accreting and synchronized long period radio transient WD-M dwarf binary ILT J110160.52+552119.62 which yields a radio luminosity of up to $\approx10^{28}$\,erg\,s$^{-1}$$^[$\cite{deRuiter25}$^]$. It has been proposed that long period radio transients may be an evolutionary phase linking WD pulsars like AR Sco to the more common polars\cite{schreiber21,rodriguez25}, although this link is being debated\cite{castroSegura25}. While the mechanism responsible for the radio emission of these systems remains unknown, and while RXJ0528+2838 does not show the same radio luminosity level as some other systems, they all share energy losses from the binary that remain to be explained, and these losses are substantial enough to alter binary evolution channels. Although all these systems have reportedly low radio luminosities, it is important to note that, at least for AR Sco, the bulk radiation is observed at infrared and radio wavelengths, yielding integrated luminosities of up to $6.3\times10^{32}$\,erg\,s$^{-1}$$^[$\cite{marsh16}$^]$. It is also important to note that while the energy supply for the radiation in AR Sco is consistent with spin-down energy of the WD, the mechanism radiating this broad component (and radio emission in particular) is still debated, with both synchrotron radiation as well as electron cyclotron maser instability\cite{marsh16,schreiber21} being considered. 

If a similar process is operating during the low states of RXJ0528+2838, similarly to what has been suggested for the synchronised and non-accreting ILT J110160.52+552119.62 system, it could provide enough energy to inflate the observed bow shock. Under this scenario the WD in RXJ0528+2838 may have directly evolved from a non-accreting synchronised system similar to ILT J110160.52+552119.6, which in turn evolved from systems akin to AR Sco where the WD surface magnetic field had been greatly enhanced, possibly through crystallisation of the WD core\cite{schreiber21}. Interestingly this scenario may also help explain why the strongest magnetic fields of $\approx10^9$G are only found in isolated WDs\cite{ferrario15}, as interacting close binaries would promptly deplete the WD magnetic energy density. Nonetheless it is clear that RXJ0528+2838 provides a rare and compelling observational window into a potentially important, yet overlooked, energy-loss channel in compact and strongly magnetized interacting binaries.

\vspace{0.5cm}

\section*{METHODS}

\subsubsection*{RXJ0528+2838 photometry}

We investigated the orbital variability of RXJ0528+2838 using data from the \textit{Transiting Exoplanet Survey Satellite} (\textit{TESS})\cite{2015JATIS...1a4003R} covering sectors 43, 44, 45, and 71. Combining all sectors we recover an orbital period of 80.05169(2)~min with no significant changes in the orbital period when comparing individual sectors. In particular, comparing the periods and associated errors measured in sectors 43 and 71\cite{1999DSSN...13...28M}, separated by 761 days, yields an upper limit on the orbital period derivative of $\dot{P}_{\text{orb}} < 2 \times 10^{-9}$~s s$^{-1}$.

The orbital variability of RXJ0528+2838 exhibits a shape typical of a polar\cite{2022MNRAS.516.5209I} (Supplementary~Data~Fig.~\ref{fig:tess}). During sectors 43, 44, and 45, we observed a dip prior to maximum light, consistent with a partial eclipse by the accretion stream\cite{2019MNRAS.489.1044B}. This eclipse was absent during \textit{TESS} sector 71, indicating a change in the accretion geometry. This change was likely caused by a decrease in the mass transfer rate, as suggested by the decrease in system brightness between \textit{TESS} sectors 45 and 71 (Supplementary~Data~Fig.~\ref{fig:long_lc}).

In asynchronous polars, the mismatch between the white dwarf spin period and the binary orbital period leads to periodic variations in the accretion geometry, which typically manifest as a peak in the periodogram at the beat frequency\cite{1994PASP..106..209P}. However, when combining all the \textit{TESS} sectors of RXJ0528+2838, no such beat frequency is detected within the accessible frequency range, placing an upper limit on the system's asynchronization of $1-P_{orb}/P_{spin}<0.02\%$. We note that we did not attempt to remove any long term systematic trends for the \textit{TESS} data but simply subtracted the mean count rate from each sector before combining all sectors.

We investigated the long-term variability of RXJ0528+2838 using photometric observations. We employed data from the All-Sky Automated Survey for Supernovae (ASAS-SN)\cite{2014ApJ...788...48S,2017PASP..129j4502K}, the Asteroid Terrestrial-impact Last Alert System (ATLAS)\cite{2018PASP..130f4505T,2018AJ....156..241H}, and the Wide-field Infrared Survey Explorer (\textit{WISE})\cite{2010AJ....140.1868W,2011ApJ...731...53M}. The \textit{WISE} observations were retrieved using the \texttt{wise\_light\_curves} Python module\cite{2020MNRAS.493.2271H}. The full light curve is presented in Supplementary~Data~Fig.~\ref{fig:long_lc}.

RXJ0528+2838 exhibited gradual long-term variability, with a broad maximum around MJD$\simeq$59300. This variability was most pronounced in the $c$ and the $o$ filters and weaker in $V/g$ filters. Since the $V/g$ band lies between the $c$ and $o$ filters in wavelength, the observed colour trends cannot be attributed to variable reddening and instead point to changes in the mass transfer rate as the primary driver of the long-term variability.

In addition to the gradual changes, RXJ0528+2838 underwent two low-accretion states, a common feature among polars\cite{2008A&A...481..433W}. While the onset of the first low state was not observed directly, it must have occurred sometime after MJD$\simeq$60420. The egress began around MJD$\simeq$60570 and lasted approximately ten days, consistent with the rapid transitions typically seen in polars. The second low state began around MJD$\simeq$60682, with the transition lasting between 3 and 8 days. A gradual recovery toward the high state appeared to begin immediately, but as of the most recent observation (MJD=60788), the system had not yet returned to the high state.

\subsubsection*{MUSE observations}

We observed RXJ0528+2838 using the Multi Unit Spectroscopic Explorer (MUSE)\cite{2010SPIE.7735E..08B} at the Very Large Telescope (VLT). The observations were made in the Wide Field Mode (WFM), which resulted in a  1x1~arcmin field of view with  0.2~arcsec sampling. The field of view was selected in order to cover the entirety of the nebula around the observed IGAPS image. The observations were carried out using the wavelength range 4750--9350 \AA\, with a spectral resolution
of R$\sim$1800--3500. In total we observed RXJ0528+2838 for 35850s spread over 33 frames. The detailed log of observations is available online under Program ID 112.25LQ.001. The observations were reduced using the MUSE Instrument Pipeline\cite{2020A&A...641A..28W} and standard MUSE procedures.  Each science exposure was accompanied by an offset sky exposure taken in an empty region near RXJ0528+2838 and reduced in the same way. For every science frame, the MUSE pipeline constructed a sky model from the corresponding offset exposure and applied it for sky subtraction. We did not apply any PCA-based sky-residual cleaning (e.g.\ ZAP) or empirical flat-fielding tools (e.g.\ CubeFix) in order to avoid the risk of suppressing extended line-dominated emission. Residuals were evaluated by inspecting line-free regions of the cube and measuring background statistics in blank apertures.

The fluxes and radial velocities of emission lines were measured fitting a Gaussian function through the \texttt{pyspeckit}\cite{2022AJ....163..291G,2011ascl.soft09001G} package. We have attempted double-Gaussian fits but could not resolve two components. The emission line fluxes across the nebula, normalized to the peak H$\alpha$ flux, are presented in Fig.~\ref{fig:lines}. Moreover, we extracted the spectra of RXJ0528+2838 itself from each individual observation.  

To mitigate slicer/IFU systematics, we rotated the observation by 90$^{\circ}$ between each frame. We combined all of the exposures using the MUSE Instrument Pipeline. We reduced and combined the two angle sets both together and separately, and the detected bow-shock features show consistent morphology and centroid in each instrument orientation. This rules out fixed-pattern artifacts associated with a single orientation. We performed a star detection on a white-light image to identify sources that could influence the detected morphology by contamination. We find that contamination from RXJ0528+2838 itself prevents the detection of {\sc [N~II]}, {\sc [O~III]}, and {\sc [S~II]} emission lines from the bow shock, and that H$\alpha$ emission from RXJ0528+2838 may blend with H$\alpha$ emission from the bow shock. Moreover, contamination from two stars southwest of RXJ0528+2838, 2MASS J05283208+2838304 and 2MASS J05283158+2838344, prevents us from measuring the bow-shock H$\alpha$, {\sc [N~II]}, and {\sc [S~II]} emission lines in their vicinity. Similarly, a star northwest of RXJ0528+2838, WISEA J052831.69+283854.1, prevents measurement of the bow-shock H$\alpha$ emission in its immediate surroundings. However, visual inspection of the stellar PSFs shows that these contaminating sources do not account for the extended low-surface-brightness features we report, and they do not significantly influence our measurements, aside from small localized artifacts visible in Fig.~\ref{fig:lines} at the positions of these stars.

The spectra of RXJ0528+2838 extracted from individual MUSE exposures reveal prominent cyclotron humps, i.e. harmonics of cyclotron frequency,  originating from cyclotron radiation produced in the post-shock region above the magnetic polar caps of the white dwarf.
The positions of these harmonics depend on the local magnetic field strength\cite{1990SSRv...54..195C}. By identifying the cyclotron humps in the spectra\cite{1999ASPC..157..127B}, we estimate a surface magnetic field strength in the range \( B \approx 42\!-\!45~\mathrm{MG} \) for the white dwarf (Supplementary~Data~Fig.~3). This is consistent with the estimate of \( B \approx 41~\mathrm{MG} \)  reported in the literature\cite{2019ASPC..518..100G}.

The emission lines in the spectra show clear variability with the orbital phase (Supplementary~Data~Fig.~4). As no donor star absorption features are evident, we measured radial velocities using the emission lines H$\beta$, He {\sc ii}~5411\AA, H$\alpha$, and He{\sc i}~6678\AA. Velocities were computed using the bisector method and averaged. We then phased the barycentric velocities using the photometric orbital period and fit a sinusoidal model. From this fit, we estimate the systemic radial velocity of RXJ0528+2838 to be \( \langle \mathrm{RV} \rangle \simeq 128~\mathrm{km\,s^{-1}} \) (Supplementary~Data~Fig.~5). While the fitted semi-amplitude of the velocities is approximately \( \sim 350~\mathrm{km\,s^{-1}} \), we caution against interpreting this directly as the white dwarf orbital motion, as the exact origin of the line-emitting region remains uncertain.

\subsubsection*{Bow shock properties}

The spatial extent of the bow shock in RXJ0528+2838 varies significantly depending on the emission line used to trace it (Fig.~\ref{fig:lines}). To measure the distance from the central system to the bow-shock apex, we fitted a second-order Taylor expansion to the bow profile around the apex, providing a simple parabolic approximation to the local bow shape\cite{2018MNRAS.477.2431T}. The uncertainty was estimated from the fit errors.
 The measured distance from the central system to the bow shock apex is largest in H$\alpha$ at 16.69$\pm$0.05~arcsec, followed by 10.50$\pm$0.08~arcsec in the {\sc [N~II]} lines, and smallest in the {\sc [O~III]} lines at 6.25$\pm$0.10~arcsec. At the distance of RXJ0528+2838 (224.2~pc$^[$\cite{2021AJ....161..147B}$^]$), this corresponds to bow shock sizes of 0.018~pc (3800~AU), 0.011~pc (2400~AU), and 0.007~pc (1400~AU), respectively.

While the overall morphology of the bow shock appears roughly symmetrical across most emission lines, its surface brightness distribution is noticeably asymmetric, with enhanced emission toward the South and South-West. The morphology in the {\sc [S~II]} lines, however, departs markedly from a typical bow shock shape. This may indicate that the physical conditions required for efficient {\sc [S~II]} emission, such as specific densities, temperatures, or ionization states, are only met in localized regions of the structure. However, a detailed interpretation of the bow shock structure is limited by the absence of line ratios sensitive to electron temperature. Furthermore, the emission line widths observed in the MUSE spectra are constrained by the relatively low spectral resolution. Higher-resolution spectroscopy may reveal intrinsic variations in line widths, potentially identifying the physical origin of the emission - whether from the shock front itself or the outflow from the CV.

The full extent of the {\sc [N~II]} tail trailing RXJ0528+2838 is not captured within the MUSE field of view (Fig.~\ref{fig:lines}), but we estimate a conservative lower limit on its projected length of 57~arcsec. The RXJ0528+2838 proper motion $\mu_{\alpha} = 39.701 \pm 0.061$~mas yr$^{-1}$ and $\mu_{\delta} = -41.6621 \pm 0.045$~mas yr$^{-1}$$^[$\cite{2021A&A...649A...1G}$^]$ corrected for the local standard of rest, together with the {\sc [N~II]} tail length, implies that the bow shock must have persisted for at least the last $\sim$1000~years.

The smooth morphology of the bow shock supports a scenario in which energy is being continuously supplied by RXJ0528+2838. Alternatively, a sequence of discrete mass-loss episodes could reproduce a similar structure, provided they occur frequently enough to remain unresolved in the available data.  The full width at half maximum  of the point spread function in the combined MUSE image is 4.3~arcsec. Given the proper motion of RXJ0528+2838, this angular resolution translates into a temporal smoothing scale of approximately 75~years. Thus, any outflow events separated by less than this timescale would be observationally indistinct, allowing for the possibility of recurrent episodic activity as the driver of the bow shock.

To constrain the physical conditions within the nebula, we used the flux ratio of the {\sc [S~II]} doublet as a diagnostic of the electron density. The observed ratio appears nearly constant across the extent of the bow shock (Supplementary~Data~Fig.~\ref{fig:sii_ratio}), indicating a uniform density throughout the structure.  We measured a mean ratio of {\sc [S~II]}~6716/6731=1.408$\pm$0.002 across the bow shock. Assuming the nebula temperature of 10,000~K this implies the electron density in the bow shock to be n$_e=42.3^{+2.9}_{-1.8}$~cm$^{-3}$. Adopting a circular radius of 0.018~pc 
we determined the mass of the bow shock to be  $4.17^{+0.29}_{-0.18}\times 10^{-5}$M$_\odot$. We note That the inferred mass is sensitive to the assumed electron temperature and reaches its maximum value of about $4\times10^{-5}$ M$\odot$ for temperatures in the range 5,000–10,000 K. At higher temperatures the mass would decrease by more than a factor of four between 10,000~K and 20,000~K, while at lower temperatures the decline is even steeper.

To estimate the strength of the presumed outflow responsible for the formation of the RXJ0528+2838 bow shock, we assume that the entire mass of the nebula must have been supplied within the time it took the system to traverse a distance equal to the bow shock current radius. Using the previously derived bow shock mass, radius, and the system proper motion, we estimate an upper limit on the mass-loss rate of $\dot{M}_o < 1.2 \times 10^{-7} M_{\odot}$/yr. This estimate does not account for the contribution of swept-up interstellar material and thus represents a conservative upper bound on the true outflow rate.

To further investigate the physical conditions within the emission-line region, we constructed diagnostic diagrams based on standard optical emission line ratios\cite{2013MNRAS.431..279S}. We applied a kernel density estimate to the full distribution of line ratios measured across the bow shock, revealing that the excitation conditions are broadly consistent with those found in supernova remnants (Supplementary~Data~Fig.~\ref{fig:diagrams}). While the nebula surrounding RXJ0528+2838 is not itself a supernova remnant, this similarity suggests the presence of shock excitation. This finding effectively rules out scenarios in which the observed nebulosity arises from a coincident H~II region or from a planetary nebula ejected during the formation of the current white dwarf.

The different bow shock sizes observed in various emission lines suggest a high stratification of the bow shock. Since the density across the bow shock appears to be uniform (Supplementary~Data~Fig.~\ref{fig:sii_ratio}), the stratification is likely due to a heterogeneous temperature distribution. Moreover, the stratification is visible in the radial velocities of different emission lines (Supplementary~Data~Fig.~\ref{fig:rvs}). Namely, the mean radial velocity of the H$\alpha$ line, 27.84$\pm$0.14~km s$^{-1}$, is the highest, being closest to the RXJ0528+2838 system putative velocity of $\simeq128$~km s$^{-1}$. The mean radial velocities of the {\sc [S~II]} and {\sc [N~II]} lines were measured to be 12.92$\pm$0.23~km~s$^{-1}$ and 11.53$\pm$0.09~km~s$^{-1}$, respectively. Their radial velocities are closest to the expected radial velocity of the interstellar medium at the position and distance of RXJ0528+2838, which is approximately +8~km~s$^{-1}$$^[$\cite{2015ApJS..216...29B}$^]$.

This, together with the extended {\sc [N~II]} emission tail, may suggest that the {\sc [S~II]} and {\sc [N~II]} emission originates from material left behind the bow shock after its interaction with the interstellar medium. The mean radial velocity of the {\sc [O~III]} emission is intermediate, at 21.12$\pm$0.85~km~s$^{-1}$. Interestingly, the radial velocities of the {\sc [N~II]}, {\sc [O~III]}, and {\sc [S~II]} emission lines remain uniform across the bow shock, despite the large variation in their surface brightness. In contrast, the H$\alpha$ line shows significant radial velocity variations, reaching its highest values near the outer edges of the bow shock. This behaviour is consistent with the broader range of physical conditions under which H$\alpha$ can be emitted, compared to the more restrictive conditions required for forbidden lines.

A comparison of the bow-shock properties of RXJ0528+2838 with those of the best-studied bow shocks around cataclysmic variables is presented in Supplementary~Data~Table~\ref{exttab:bowshock-properties}. Notably, the RXJ0528+2838 bow shock requires an energy input comparable to that of systems where the outflow is thought to be powered by an accretion-disk wind, but its mass and size differ significantly.

\begin{supptable*}[t]
\centering
\caption{\textbf{Properties of bow-shock nebulae around cataclysmic variables.} The properties include the space velocity relative to the local interstellar medium ($v_{\mathrm{ISM}}$), bow-shock size, mechanical luminosity required to power the bow shock ($L_{b}$), and total bow-shock mass.}
\label{exttab:bowshock-properties}
\begin{tabular}{lccc}
\hline
Property & RXJ0528+2838 & BZ Cam & V341 Ara \\
\hline
$v_{ISM}$ [km s$^{-1}$] 
  & $142$ 
  & $125$ $^[$\cite{2001A&A...376.1031G}$^]$ 
  & $76 \pm 1$$^[$\cite{2021MNRAS.501.1951C}$^]$ \\

Size [pc] 
  & $0.018$ 
  & $1.2$ {$^[$\cite{2001A&A...376.1031G}$^]$ }
  & $(5.5 \pm 0.3)\times 10^{-3}$$^[$\cite{2021MNRAS.501.1951C}$^]$ \\

$L_{b}$ [erg s$^{-1}$] 
  & $8.2 \times 10^{32}$ 
  & $1.1 \times 10^{32\pm1}$$^[$\cite{1992ApJ...393..217H}$^]$ 
  & $(1.2 \pm 0.2)\times 10^{32}$$^[$\cite{2021MNRAS.501.1951C}$^]$ \\

Mass [$M_{\odot}$] 
  & $(4.17^{+0.29}_{-0.18}) \times 10^{-5}$ 
  & $(2.4 \pm 0.4)\times 10^{-3}$$^[$\cite{1992ApJ...393..217H}$^]$ 
  & $5 \times 10^{-3}$$^{(*)}$$^[$\cite{2008PhDT.......109F}$^]$ \\
\hline
\end{tabular}

\begin{flushleft}\footnotesize
$^{(*)}$ Total mass includes the bow shock together with the larger H$\alpha$ nebula.
\end{flushleft}
\end{supptable*}

\subsubsection*{{\it XMM-Newton} and  {\it Swift} observations}

RXJ0528+2838 was discovered as an X-ray source by {\it ROSAT} and identified as 1RXSJ\,052832.5+283824.  We requested two short (1.7\,ks and 1.0\,ks separated by $\sim$5\,hrs) pointings with the XRT instrument on-board the {\it Neil Gehrels Swift} satellite\cite{2005SSRv..120..165B}, performed on February 24, 2024  (ObsId: 00016536001, PI:Ilkiewicz) (see Supplementary~Data~Tab.~\ref{tab:tabxrayobs}). Processing of the data, light curve and spectrum extraction were performed through the UKSSDC on-line product generator\cite{2020ApJS..247...54E}.
The X-ray source was detected with a net average count rate of 0.104$\pm$0.07\,$\rm cts\,s^{-1}$ in
the 0.3-10\,keV range. It displayed a large variability (more than a factor
of two) reminiscent of orbital modulation of polar-type CVs, but the poor coverage could not allow to infer the orbital period. The {\it Swift}/XRT spectrum in the 0.3-10\,keV range was fit using Cash statistics with an absorbed power law model {\sc tbabs*pow} within the  {\sc xspec} package\cite{arnaud96}
where {\sc tbabs} accounts for the ISM absorption with abundances from\cite{wilms00}, giving a power law index $\alpha = 1.18_{-0.23}^{+0.25}$ and hydrogen column density $\rm N_H = 1.14_{-0.89}^{+1.13}\times 10^{21}\,cm^{-2}$ (c-stat/d.o.f.=151/193). The unabsorbed flux in the 0.3-10\,keV range results to be $\rm 9.5\pm1.4\times10^{-12}\,erg \,s^{-1 } cm^{-2}$. The HI column density is much lower than that of the ISM in the direction of the source ($\rm N_H = 3.7\times 10^{21}\,cm^{-2}$)$^[$\cite{HI4PI16}$^]$ consistent with the distance of 224\,pc. A spectral fit using an absorbed optically thin plasma {\sc apec} was also attempted providing an unconstrained temperature $\rm kT >$ 12\,keV.

We observed RXJ0528+2838 with  {\it XMM-Newton} on 2025 February 23 via a DDT request (ObsID: 0954191301) with the EPIC-pn camera\cite{Struder01} and with the EPIC-MOS1 and MOS2 cameras\cite{Turner01} in the Full Frame mode and with the thin optical blocking filters for 30.1\,ks (EPIC-pn) and 34.9\,ks (EPIC-MOS1,2). The Optical monitor (OM)\cite{Mason01} was set in fast window mode using the B (3975-5025 \AA) filter. Six OM science windows of $\sim$4400\,s each were acquired sequentially totalling an exposure of of 26.4\,ks. The log of the {\it XMM-Newton} observations is reported in Supplementary~Data~Tab.~\ref{tab:tabxrayobs}.

\begin{supptable*}
\centering
 \begin{minipage}{140mm}
\caption{\textbf{The X-ray observing log of RXJ0528+2838.}} The list of observations carried out with {\it Swift}  and {\it XMM-Newton}.
\label{tab:tabxrayobs}
\begin{tabular}{lccccc} 
\hline
Telescope & Instrument & Date      & UT start (hrs) & $\rm T_{expo}$ & Average rate (cts\,s$^{-1}$)$^a$\\
\hline
{\it Swift} & XRT      & 2024-02-24  &  07:55:56 & 1.67\,ks & 0.09$\pm$0.04\\
            &           &            &  12:50:56 & 1.07\,ks & 0.11$\pm$0.04\\
\hline
{\it XMM-Newton} & EPIC-pn  & 2025-02-23 & 04:49:28 & 30.1\,ks & 0.44$\pm$0.08\\
     & EPIC-MOS1 &           & 04:26:21 & 34.9\,ks & 0.13$\pm$0.04\\
     & EPIC-MOS2 &           & 04:26:40 & 34.9\,ks & 0.13$\pm$0.04\\
     & OM-B      &           & 05:53:06 & 26.4\,ks & 4.37$\pm$0.43 (17.75$\pm$0.12)\\
\hline
\end{tabular}
$^a$ Average net count rates in the 0.3-10\,keV range for XRT and in the 0.2-12\,keV range for EPIC cameras. For the OM, the average rate and the Vega magnitude in parenthesis are reported.
\end{minipage}
\end{supptable*}

The data were processed and analysed using the {\it XMM-Newton} Science Analysis Software ({\sc SAS} v.20.0) with the latest calibration files. The photon arrival times from EPIC cameras were reported to the solar system barycentre using the JPL DE405 ephemeris and the nominal {\em Gaia} DR3 position\cite{gaiadr3}  of RXJ0528+2838. For the EPIC cameras, we extracted events using a 15" radius circular region centred on the source and using a background region of the same size located on the same CCD chip.
To improve the S/N the data were filtered by selecting pattern pixel events up to double with zero-quality flag for the EPIC-pn data and up to quadrupole for the EPIC-MOS data. The presence of background flaring activity at the beginning of the pointing
was found using high energy ($>$10\,keV) coverage, which resulted in no effects on the light curve analysis. However, conservatively, the high background period was removed from the EPIC event files when extracting the X-ray spectra, resulting  in a reduction of effective exposure times of 26.5\,ks and 29.7\,ks for the EPIC pn and MOS cameras, respectively. 
We produced background subtracted light curves using the {\sc epiclccorr} task in the total 0.2-12\,keV range with 10\,s binning and in three energy bands covering 0.2-2\,keV, 2-4\,keV and 4-12\,keV ranges with 200\,s bin time.

To extract the EPIC 0.3-10\,keV spectra we previously filtered the events using the task {\sc gtigen} with  good time intervals determined by selecting low background epochs.  We generated response matrix and ancillary files using {\sc rmfgen} and {\sc arfgen}, respectively. Finally we used {\sc specgroup} routine to rebin the spectra with a minimum of 30 counts in each bin for fitting purposes.

The optical B-band data from the OM  were processed using the task {\sc omfchain} and generating background subtracted light curves with bin time of 100s. Each fast window data was inspected against target centring ensuring that no drift of the telescope occurred during the observation. The light curves were also corrected to the solar system baricentre with the same procedure as the EPIC data. We converted the B-band count rates  into Vega magnitudes following the count rate magnitude conversion reported at the {\it XMM-Newton} Science Operation Centre.

The 80\,min orbital modulation is clearly visible in all instruments as well as in  all the three selected bands (Supplementary~Data~Fig.~\ref{fig:xmm_final}). The power spectra of the EPIC-pn and OM B display the orbital fundamental frequency ($\Omega$) and the first two harmonics (2$\Omega$ and 3$\Omega$) well above the 95$\%$ and 99$\%$ single trial confidence levels as well as above the global 95$\%$ and 99$\%$ confidence levels, with the first harmonic (2$\Omega$) being more prominent than the fundamental (Supplementary~Data~Fig.~\ref{fig:xmm_final}). 

Period search using the three frequencies $\Omega$, 2$\Omega$ and 3$\Omega$, linked together gives
an X-ray period of  4800.1 $\pm$ 8.6\,s and an optical period of 4799.4$\pm$5.3\,s (errors at 1$\sigma$ level), both consistent with the better determined orbital period from TESS 4803.101$\pm$0.001\,s and by other authors\cite{2016AstBu..71...95B}. 

Folded light curves at the 4803\,s TESS period in the X-ray (0.2-12\,keV) and B-band ranges (see Supplementary~Data~Fig.~\ref{fig:folded_xmm}, left panel) display a main wide stronger peak and a secondary weaker peak offset by 0.5 in orbital phase indicating the contribution of two accreting poles. The modulation
amplitude is more than a factor of two in both bands. Energy resolved folded light curves in 0.2-2\,keV, 2-4\,keV, 4-12\,keV also display the same behaviour with similar amplitudes in all bands.
Hardness ratios indeed do not show energy dependent variability (see Supplementary~Data~Fig.~\ref{fig:folded_xmm}, right panel)
indicating that the orbital modulation is due to changes in the viewed emitting area and not to local photoelectric absorption.

The X-ray spectra of RXJ0528+2838 from the EPIC-pn and MOS1,2 instruments were simultaneously fitted with several models. An absorbed power law did not provide an satisfactory fit giving a large excess in the residuals at the iron line complex. Thus the spectra were fitted with an absorbed thermal plasma {\sc const*tbabs*apec} where the constant accounts for normalization among the three instruments. An excess is still found requiring the addition of a second thermal component, thus {\sc const*tbabs*(apec+apec)}, giving
an hydrogen column density $\rm N_H= 1.29\pm0.10\times 10^{+21}\,cm^{-2}$ and plasma temperatures 
$\rm kT_c = 1.23_{-0.23}^{+0.13}\,keV$, $\rm kT_h = 8.27_{-0.71}^{+1.05}\,keV$ and a metal abundance $\rm A_Z= 1.68_{-0.33}^{+0.35}$ ($\rm \chi^2/d.o.f.=388.37/364 = 1.09$). A similar fit is found using a multi-temperature model  {\sc const*tbabs*cemekl} with a power-law emissivity index $\alpha=1.61_{-0.24}^{+0.31}$
and maximum temperature $\rm kT_{max}=14_{-2}^{+3}\,keV$ and abundance $\rm A_Z=1.8\pm0.4$ ($\rm \chi^2/d.o.f.=388.0/364 = 1.09$) (Supplementary~Data~Fig.~\ref{fig:xmm_spec}). Worth noticing that fixing the metal abundance to solar value yields an excess at 6.7\,keV but not at the 6.4\,keV iron $K_ {\alpha}$ fluorescent line. Adding a Gaussian to the latter we derived 
an upper limit of $<$40\,eV to the equivalent width of this line. The high  abundance indicates a slightly supra-solar value, which is not uncommon in CV systems. Unfortunately the data from the Reflection Grating instruments (RGS1,2) onboard {\it XMM-Newton} are not of sufficient quality to better constrain the metal abundance.  Both models provide evidence of lack of intrinsic, local absorption, indicating that the orbital modulation is due to changes of the emitting poles as the WD rotates. The source X-ray unabsorbed flux in the 0.3-10\,keV range is $\rm 1.7\pm0.1\times10^{-12}\,erg\,cm^{-2}\,s^{-1}$ giving a luminosity of $\rm \sim 1.0\times^{31}\,erg\,s^{-1}$. When comparing this flux with that from {\it Swift} the source was fainter by an order of magnitude during the 2025 observation, which is consistent with the optical long-term light curve shown in (Supplementary~Data~Fig.~\ref{fig:long_lc}).

The bolometric luminosity of the X-ray emitting component results in $\rm1.4\times10^{31}\,erg\,s^{-1}$. We note that this does not account for the entirety of the accretion luminosity. The cyclotron flux, emitted in the optical band, cannot be determined from the OM data since it provides coverage in only a single narrow optical band. It also cannot be reliably estimated from the MUSE spectra, which are not flux-calibrated and were obtained during a higher accretion state than the XMM observations. 

We also searched for a possible extended X-ray emission around RXJ0528+2838 on scales from $2^{\prime\prime} - 200^{\prime\prime}$ in search for an X-ray counterpart to the bow shock. The field is crowded of X-ray sources, and the closest one was detected at $\sim240^{\prime\prime}$ from our target. 

We extracted the surface brightness profiles of the XMM–Newton observation in four different energy bands (0.2--12\,keV, 0.2–2 keV, 2–5 keV and 5–12 keV) out to a radius of $200^{\prime\prime}$ (in increments of $2^{\prime\prime}$). To search for a diffuse emission around the source we then fitted the surface brightness with an analytical King PSF model, that has been proven to provide a good description of the EPIC PSF\cite{Tiengo2010}, plus a constant background term $b_0$. 
We fitted the function: $$SB = \frac{a_0}{(1+(r/r_c)^2)^{\alpha} + b_0}$$
where r is the distance from the source position, $a_0$, $r_c$, $\alpha$ are free parameters, and $b_0$ was fixed at the background level measured far from the source in the same chip. We did not find any radial excess of counts in any of the four energy ranges between $2^{\prime\prime} - 200^{\prime\prime}$. We derive a 3$\sigma$ upper limit on the flux of the X-ray diffuse emission of $< 4\times10^{-14}$\,erg\,s$^{-1}$\,cm$^{-2}$.

\subsubsection*{Radio observations}
We targeted RXJ0528+2838 with the MeerKAT array\cite{jonas16} under a Director’s Discretionary Time programme. 
The first epoch commenced on 2025-02-02 at 16:28:58.996~UTC. Observations were taken in L band, centred at 1.28~GHz, with a total processed bandwidth of 0.86~GHz spanning 856--1712~MHz. The correlator setup delivered 4096 frequency channels and an integration time of 8~s per visibility. After an initial round of RFI flagging, the data were spectrally averaged to 1024 channels for subsequent processing. Between 62 out of 64 antennas contributed to the visibilities, with baselines extending out to 7.698~km.

The main observing block (ID: 1738513549) lasted 120~min and was partitioned as follows: 90~min on RXJ0528+2838 (three 30~min scans), 10~min on the primary flux/bandpass calibrator J0408$-$6545, 10~min on the polarisation calibrator J0521+1638, and 5~min~40~s on the nearby phase calibrator J2011$-$0644 split into four short scans bracketing the target scans. Subsequent (short) tracks adopted a standard cadence of 15~min on target, bookended by two 2~min scans on the secondary calibrator and a 10~min scan on a primary calibrator.

Data reduction employed a set of dedicated \textsc{Python} scripts (\textsc{OxKAT}\cite{oxcat}) to carry out semi-automated calibration and imaging. Within \textsc{CASA}\cite{CASAteam2022} we first removed the outer 100 channels at each band edge, excised autocorrelations and zero-amplitude visibilities, and performed additional time/frequency RFI flagging. Flux-density scale, bandpass, and instrumental delay solutions were derived from the primary calibrator, while complex gains and residual delays were obtained from the secondary calibrator. A spectral model of the phase calibrator was constructed by temporarily binning the band into eight equal spectral windows tied to the primary-calibrator scale. The derived solutions were applied to the target field, which was then split to a separate measurement set, averaged to 8~s in time and by a factor of 8 in frequency for imaging, and further cleaned of residual RFI using \textsc{TRICOLOUR}\cite{tricolour}.

Imaging was performed with \textsc{WSClean}\cite{wsclean}. We first produced a wide-field image (field of view $\simeq 1.5~\mathrm{deg}^2$) and from it generated a deconvolution mask. Using this mask, we re-imaged the dataset and then applied direction-independent self-calibration with \textsc{CUBICAL}\cite{cubical}, solving for phase and delay on 32~s intervals. The resulting field image is shown in Supplementary~Data Fig.~\ref{fig:MeerKAT_field}.

No emission is detected at the catalogued position of RXJ0528+2838. The final map reaches an rms noise of $13~\mu$Jy~beam$^{-1}$, from which we adopt a conservative $3\sigma$ upper limit of $39~\mu$Jy for the source’s L-band flux density.

\section*{Acknowledgements}
~\\
 
This work has made use of data from the Asteroid Terrestrial-impact Last Alert System (ATLAS) project. The Asteroid Terrestrial-impact Last Alert System (ATLAS) project is primarily funded to search for near earth asteroids through NASA grants NN12AR55G, 80NSSC18K0284, and 80NSSC18K1575; byproducts of the NEO search include images and catalogs from the survey area. This work was partially funded by Kepler/K2 grant J1944/80NSSC19K0112 and HST GO-15889, and STFC grants ST/T000198/1 and ST/S006109/1. The ATLAS science products have been made possible through the contributions of the University of Hawaii Institute for Astronomy, the Queen’s University Belfast, the Space Telescope Science Institute, the South African Astronomical Observatory, and The Millennium Institute of Astrophysics (MAS), Chile.
This work is based on observations obtained with {\it XMM-Newton } an ESA science mission with instruments and contributions directly funded by ESA Member States and NASA.We acknowledge the use of public data from the Swift data archive. 
This work has made use of data from the European Space Agency (ESA) mission Gaia ( https://www.cosmos.esa.int/gaia ), processed by the Gaia Data Processing and Analysis Consortium (DPAC, https://www.cosmos.esa.int/web/gaia/dpac/consortium).
Funding for the DPAC is provided by national institutions, in particular the institutions participating in the Gaia MultiLateral Agreement (MLA).
This work is based on observations collected at the European Southern Observatory under ESO programme 112.25LQ.001.
This research made use of APLpy, an open-source plotting package for Python\cite{2012ascl.soft08017R}.

KI was supported by the Polish National Science Centre (NCN) grant 2024/55/D/ST9/01713. 
SS acknowledges support by the Science and Technology Facilities Council (STFC) grant ST/T000244/1 and ST/X001075/1. 
DdM acknowledges financial support from INAF (Large Research Grant ``Uncovering the optical beat of the fastest magnetised neutron stars n.16 (FANS)''.
CK acknowledge support by the Science and Technology Facilities Council grant ST/V001000/1.
NR is supported by the European Research Council (ERC) via the Consolidator Grant ``MAGNESIA'' (No. 817661) and the Proof of Concept ``DeepSpacePulse'' (No. 101189496), by the Catalan grant SGR2021-01269 (PI: Graber/Rea), the Spanish grant ID2023-153099NA-I00 (PI: Coti Zelati), and by the program Unidad de Excelencia Maria de Maeztu
CEX2020-001058-M.
DB is partially supported by a rated researcher grant (UID 132095) from the National Research Foundation.
NCS acknowledge support from the Science and Technology Facilities Council (STFC) grant ST/X001121/1.
PJG is partially supported by NRF SARChI grant 111692.


\section*{Author Contributions}
~\\
KI led the preparation of the observing proposal for both MUSE and Swift, performed the data reduction for the MUSE observations, measured the properties of the bow shock, conducted the analysis of the photometric and spectroscopic data, measured the magnetic field of the white dwarf and wrote and edited the text.
SS initiated and coordinated the project from the conception stage by designing the search for extended emission in IGAPS images, contributed to the MUSE proposal, performed the calculations on the energetics of the bow shock, wrote and edited the text, lead the \textit{TESS} proposals to obtain lightcurves, contributed to the \textit{Swift}, \textit{XMM-Newton} and MeerKAT proposals, as well as leading and suggesting possible hypotheses to interpret the result. 
DdM contributed to the preparation of \textit{XMM-Newton} and MeerKAT proposals, analysed the \textit{Swift} and \textit{XMM-Newton} data, contributed to the interpretation of the multi-wavelength data and to the editing of the text. 
SEM led the preparation of the MeerKAT proposal, led the data reduction and analysis of the radio data, and contributed to the interpretation of the multi-wavelength data. 
NR contributed to the preparation of \textit{XMM-Newton} and MeerKAT proposals, analysed the diffuse emission in the \textit{XMM-Newton} data and contributed to the interpretation of the results. AFM contributed to the MUSE proposal and data reduction. 
LTP identified the first IGAPS image containing the extended resolved structure that provided the basis for the MUSE proposal under the supervision of SS. 
CK, NCS, PJ, DB and MV all contributed to the interpretation of the results and editing of the text. 

\section*{Competing Interests}
~\\
The authors declare no competing interests.

\section*{Data availability}
~\\
The final MUSE datacube is available under DOI:10.5281/zenodo.17238003\cite{MUSE_cube}. 
The raw MUSE data can be found in the ESO archive under Program ID 112.25LQ.001.
The \textit{TESS} data can be found in the Mikulski Archive for Space Telescopes (MAST) archive by searching for TIC ID 286042374.
The \textit{XMM-Newton} can be found in the....
The \textit{Swift} data can be found in the NASA High Energy Astrophysics Science Archive Research Center under observation ID 00016536001.
The un-calibrated MeerKAT visibility data presented in this paper are publicly available in the archive of the South African Radio Astronomy Observatory at \url{https://archive.sarao.ac.za}. All other datasets are available from their respective public archives.

\section*{Materials and Correspondence}
~\\
Correspondence can be addressed to:
\\
simone.scaringi@durham.ac.uk

   \begin{suppfigure*}
   \centering
   \includegraphics[width=0.99\columnwidth]{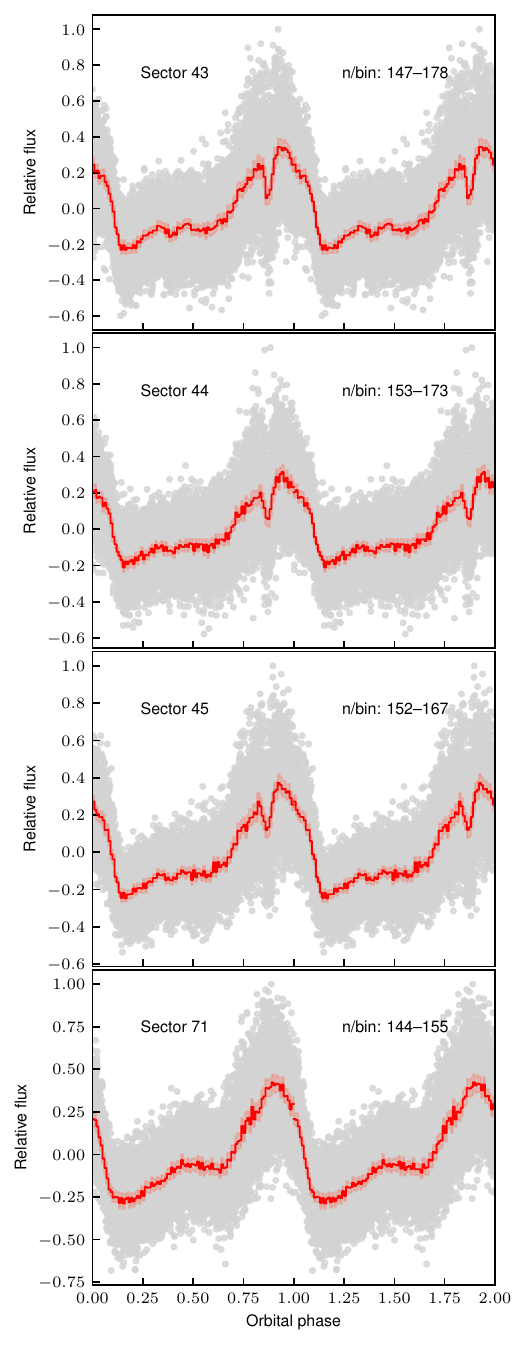}
      \caption{\textbf{Folded \textit{TESS} lightcurves of RXJ0528+2838 on the orbital period of 80.05169 minutes.} Data is from sectors 43, 44, 45, and 71 (gray points). Binned observations with a phase bin width of 0.01 are shown as red lines, with the shaded regions indicating the standard error of the mean for each bin. Each bin contained at least 140 data points. }
         \label{fig:tess}
   \end{suppfigure*}
   
   \begin{suppfigure*}
   \centering
   \includegraphics[width=0.99\columnwidth]{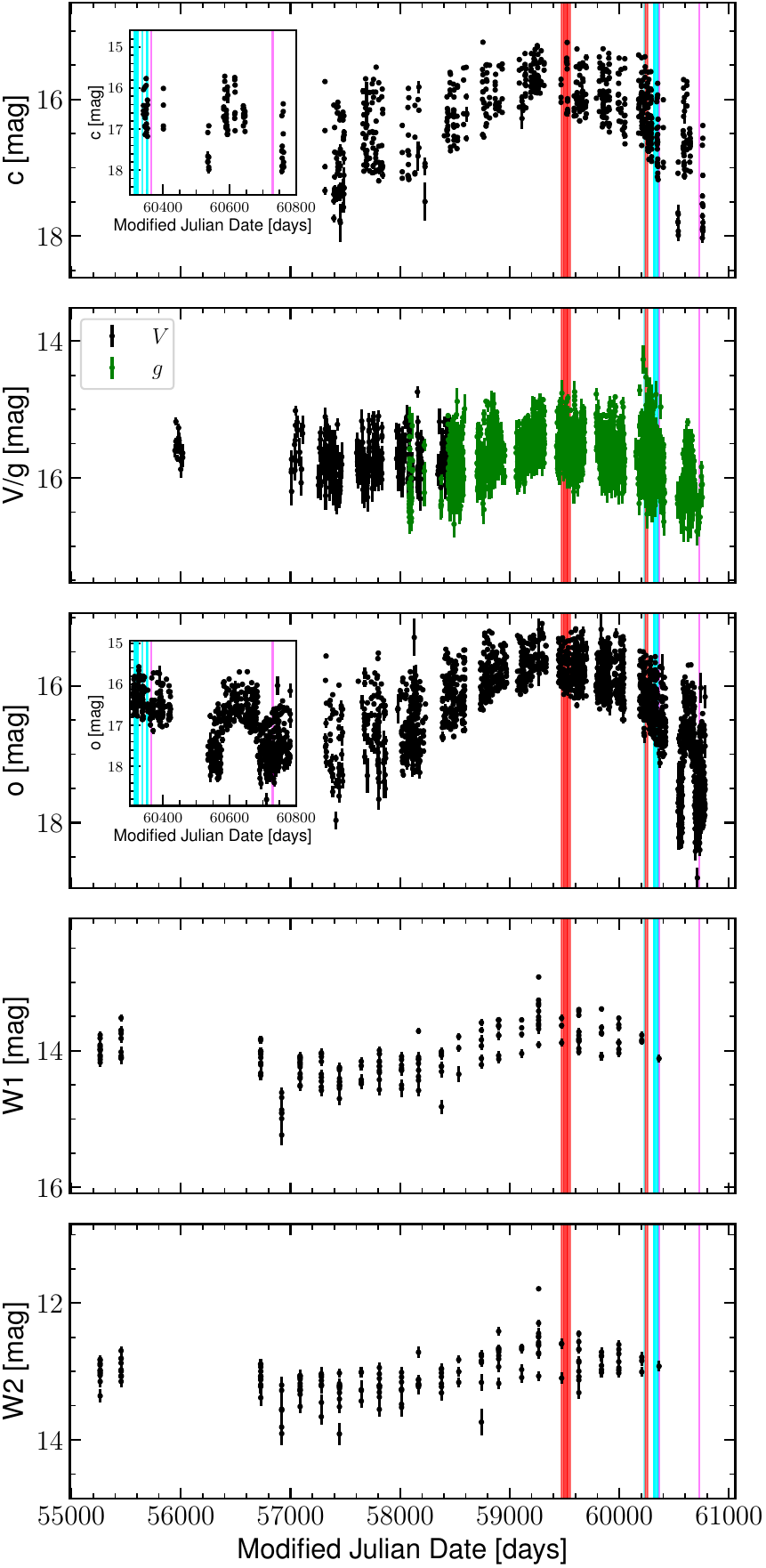}
      \caption{\textbf{Long-term photometric variability of RXJ0528+2838 across optical and near-infrared bands.} Red shaded regions indicate the \textit{TESS} observation windows, cyan vertical lines mark the times of VLT/MUSE observations, while magenta vertical lines mark the times of X-ray observations. The inset panels show zoom-ins of the relevant lightcurves around the time of the MUSE observations. All error bars display the $1\sigma$ uncertainty and are retrieved from the respective available public datasets.}
         \label{fig:long_lc}
   \end{suppfigure*}

     \begin{suppfigure*}
   \centering
   \includegraphics[width=0.99\columnwidth]{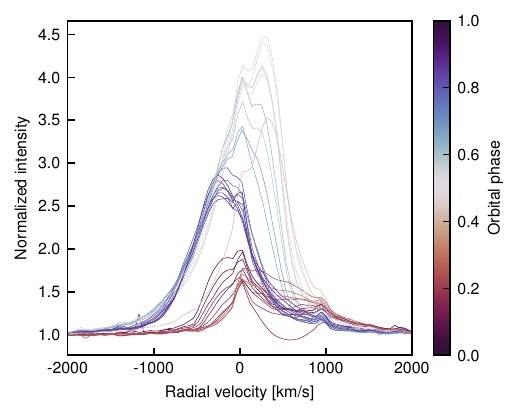}
   \includegraphics[width=0.99\columnwidth]{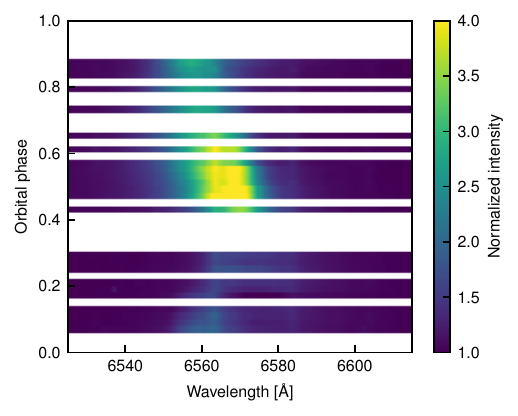}
      \caption{\textbf{Phase–resolved variability of the continuum–normalized H$\alpha$ emission line in RXJ0528+283.} Left panel: Individual spectra colour–coded by orbital phase. Right panel: Trailed spectrogram of the phase-folded spectra, with the colour scale indicating normalized intensity. The orbital ephemeris is based on photometric observations, where phase zero corresponds to the time of maximum brightness\cite{2015AstBu..70..460G}.}
         \label{fig:halpha_phased}
   \end{suppfigure*}

   \begin{suppfigure*}
   \centering
   \includegraphics[width=0.99\columnwidth]{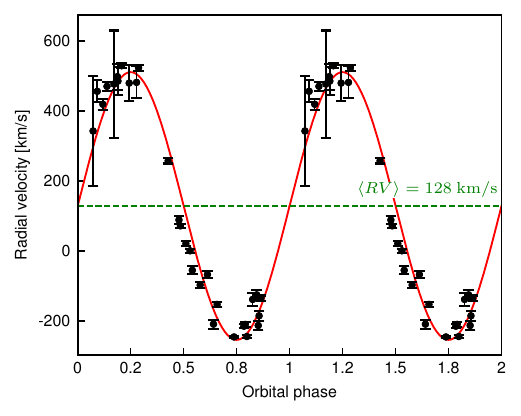}
      \caption{\textbf{Radial velocity curve of RXJ0528+2838.} Radial velocities were derived from the H$\beta$, He\,\textsc{ii}\,5411\,\AA, H$\alpha$, and He\,\textsc{i}\,6678\,\AA\ emission lines using the bisector method (black points). All error bars represent the standard error of the mean derived from the four emission lines. The red curve shows the best-fitting sinusoidal model to the 30 mean radial velocities, while the green dashed line marks the systemic velocity \( \langle \mathrm{RV} \rangle \simeq 128~\mathrm{km\,s^{-1}} \) obtained from the fit. }
         \label{fig:RVs_phased}
   \end{suppfigure*}

    \begin{suppfigure*}
   \centering
   \includegraphics[width=0.99\columnwidth]{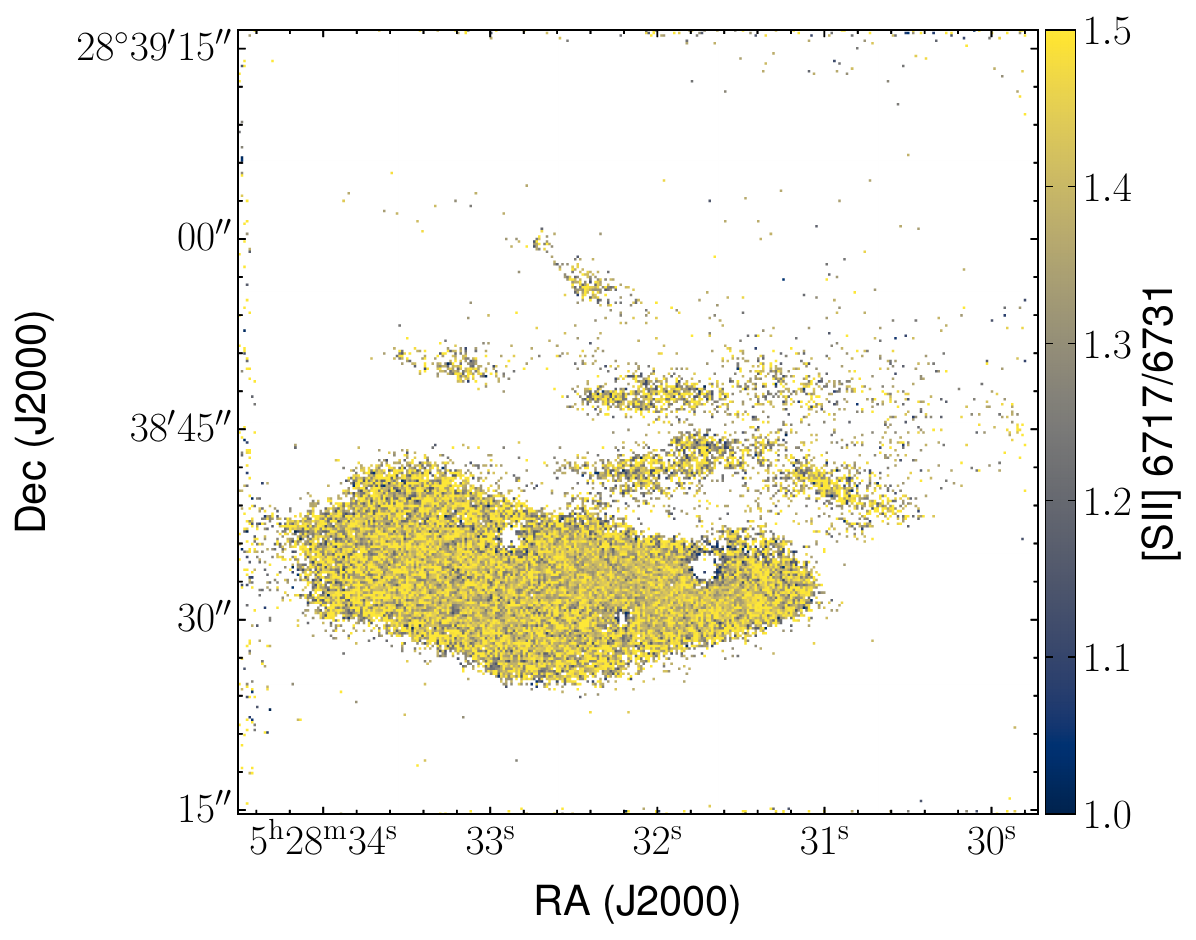}
      \caption{\textbf{{\sc [S~II]}~6716/6731 ratio measured across the bow shock in RXJ0528+2838}. The uniform distribution of the {\sc [S~II]} ratio suggests that the density across the bow shock is uniform.}
         \label{fig:sii_ratio}
   \end{suppfigure*}

   \begin{suppfigure*}
   \centering
   \includegraphics[width=0.99\columnwidth]{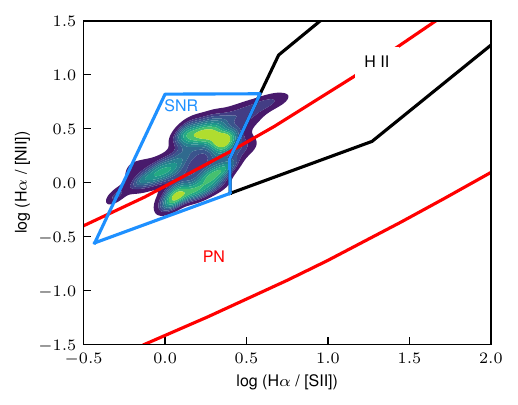}
   \includegraphics[width=0.99\columnwidth]{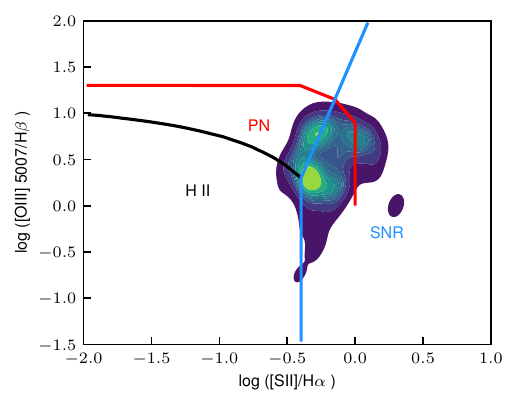}
      \caption{\textbf{Emission line diagnostic diagram for RXJ0528+2838}. The majority of the bow shock has emission lines ratios consistent with a supernova remnant, suggesting that the emission originates in a region with a shock suggesting the bow shock is a relatively new structure as opposed to a planetary nebula ejected by the current white dwarf or a chance alignment with an H~II region.}
         \label{fig:diagrams}
   \end{suppfigure*}

   \begin{suppfigure*}
   \centering
   \includegraphics[width=0.99\columnwidth]{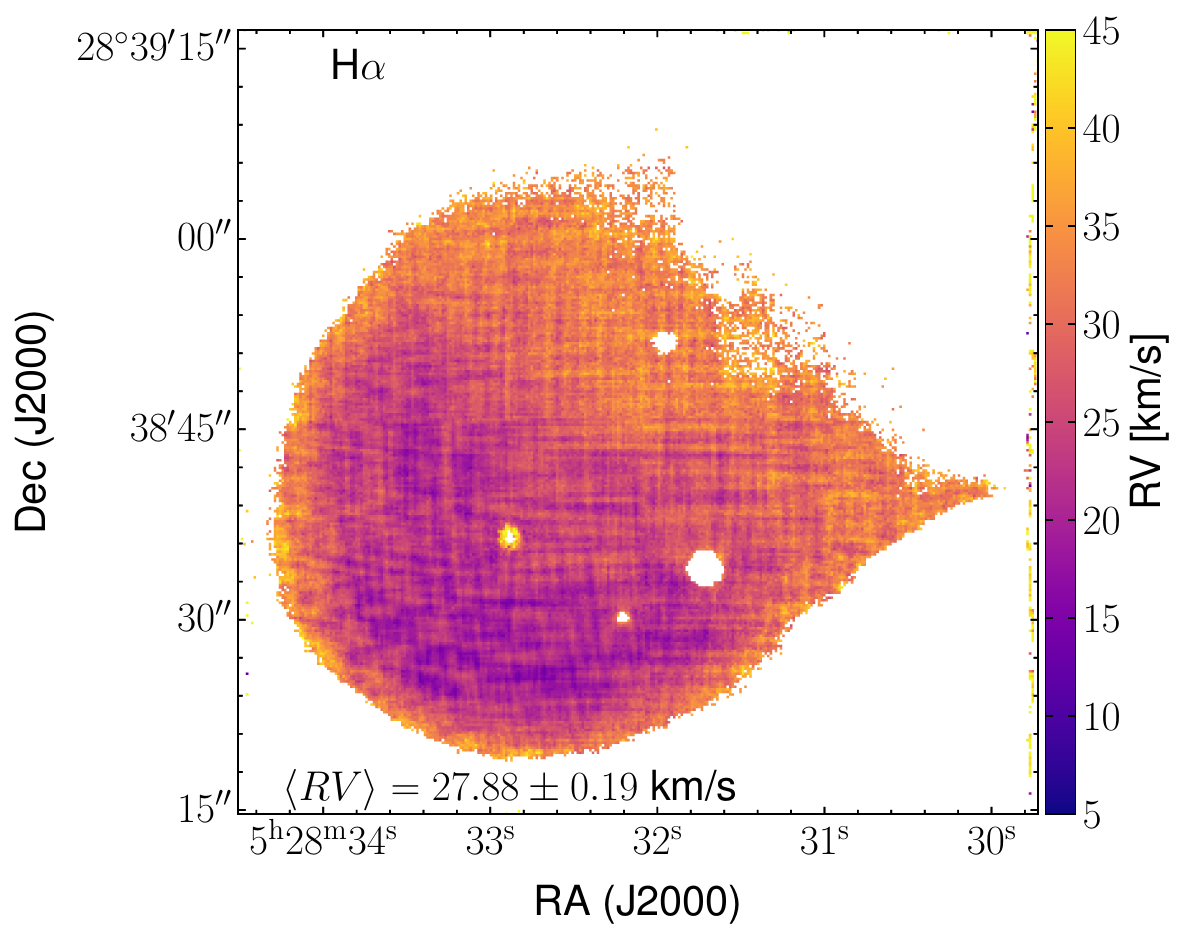}
   \includegraphics[width=0.99\columnwidth]{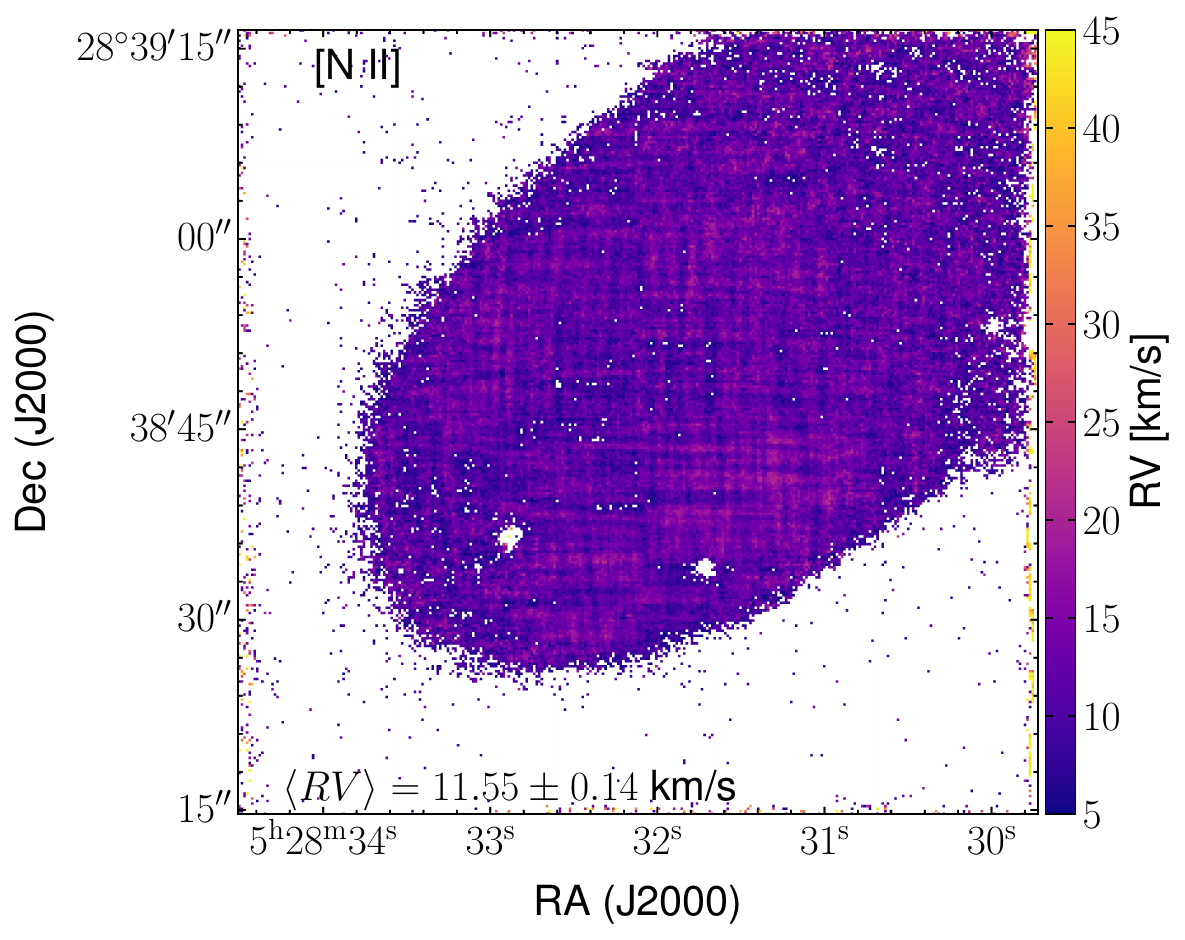}
   \includegraphics[width=0.99\columnwidth]{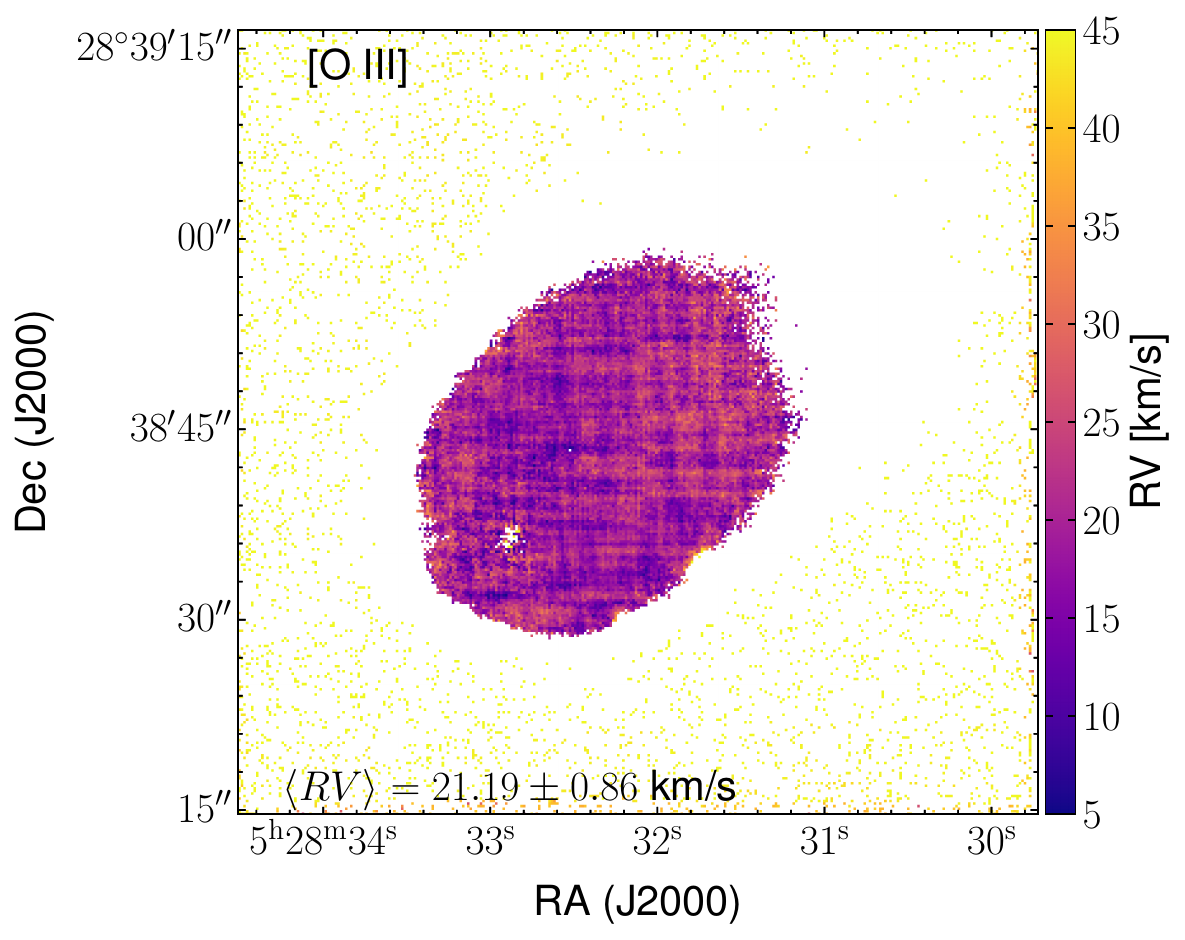}
   \includegraphics[width=0.99\columnwidth]{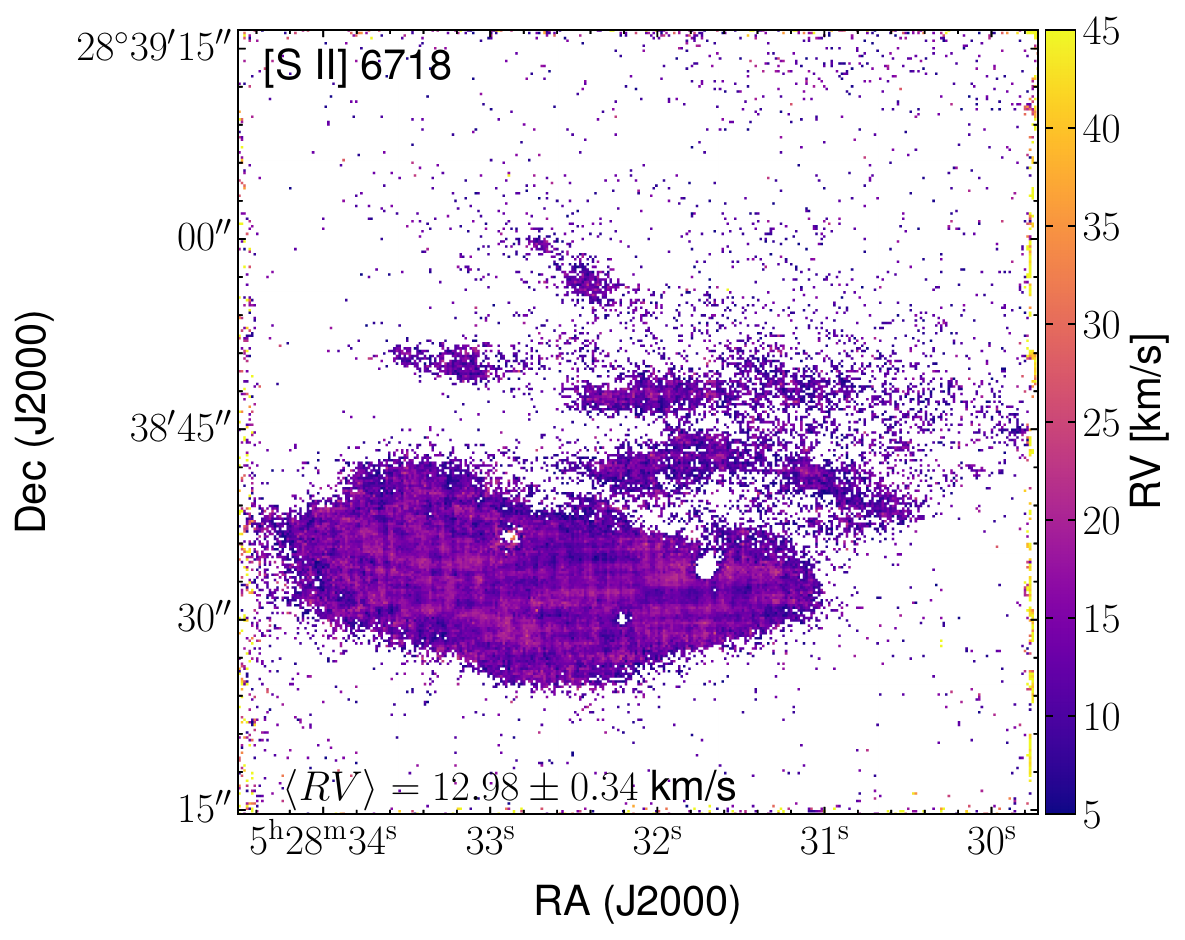}
      \caption{\textbf{Radial velocity maps of the bow shock around RXJ0528+2838.} Insets are for H$\alpha$ (top left), {\sc [N~II]} (top right), {\sc [O~III]} (bottom left), and {\sc [S~II]} (bottom right) emission lines in the bow shock.}
         \label{fig:rvs}
   \end{suppfigure*}

   \begin{suppfigure*}
   \centering

   \includegraphics[width=0.99\columnwidth]{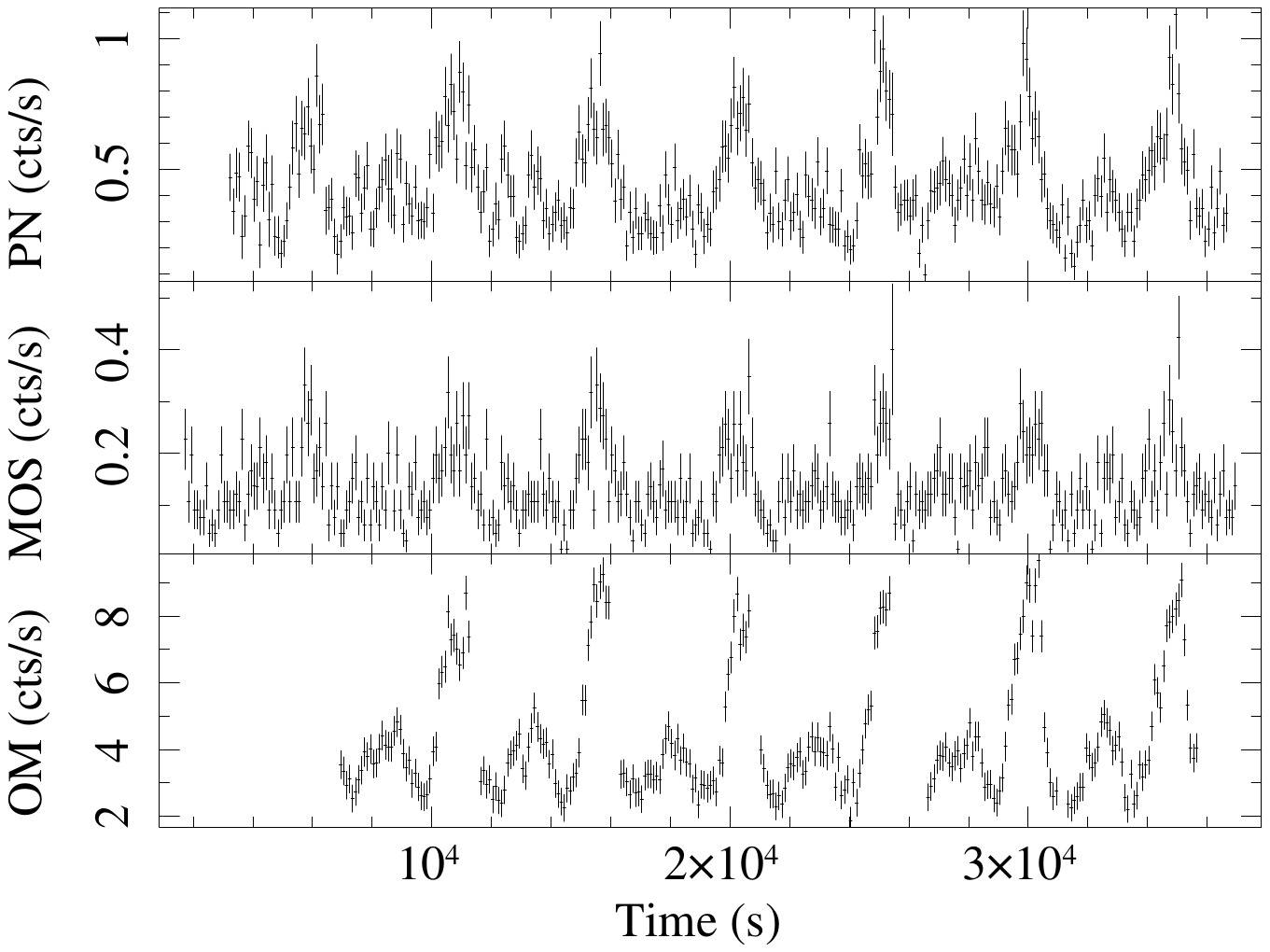}
   \includegraphics[width=0.99\columnwidth]{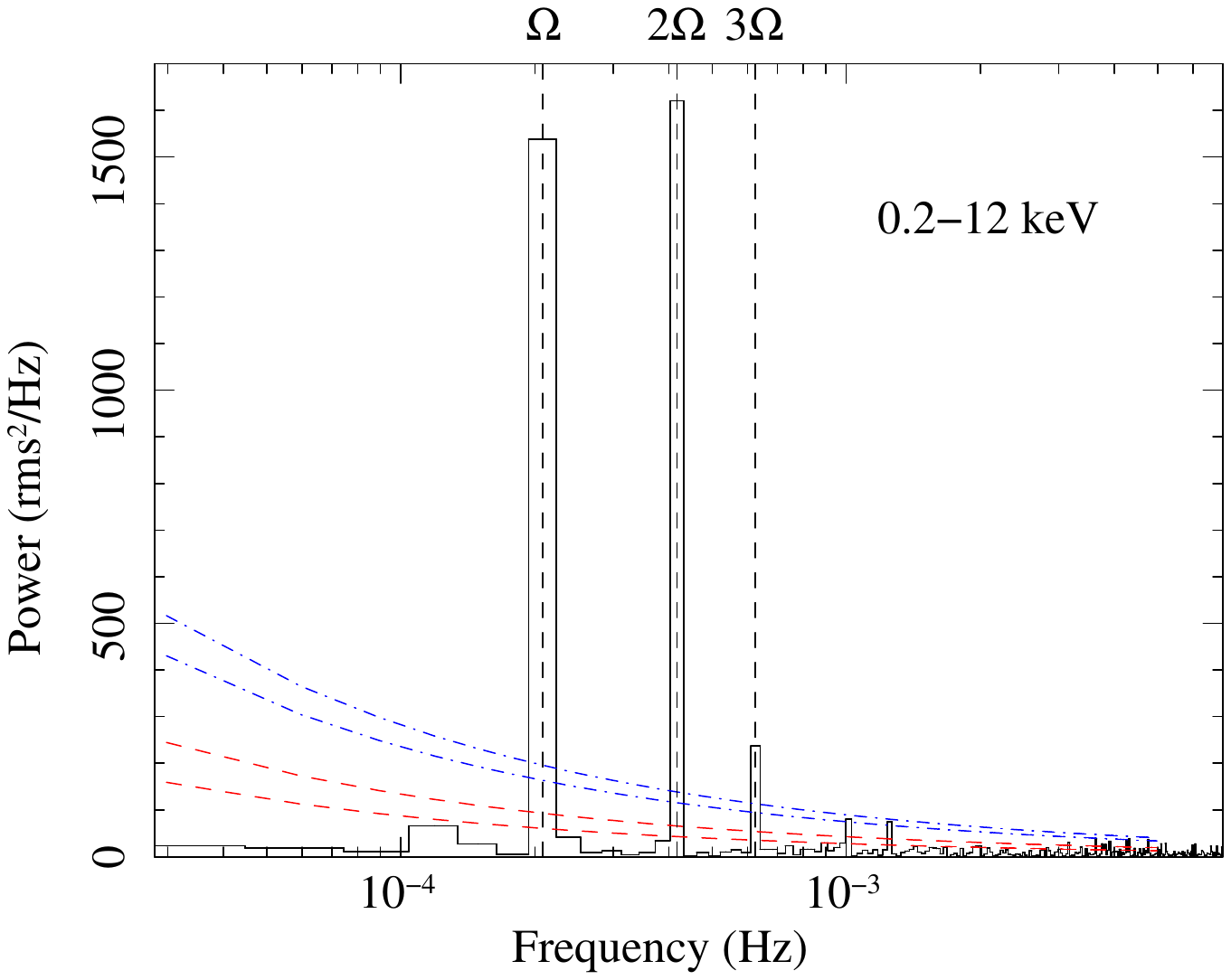}
      \caption{\textbf{X-ray lightcurves and power spectra of RXJ0528+2838.} {\it Left:} The X-ray light curves from the EPIC-pn and MOS instruments in the 0.2-12\,keV range and the optical B-band light curve binned with 100\,s intervals. All error bars display the $1\sigma$ uncertainty. {\it Right:} The power spectrum from the EPIC-pn instrument in the 0.2-12\,keV range. Single trial 95$\%$ and 99$\%$ confidence levels (red dashed lines) and the global 95$\%$ and 99$\%$ confidence levels  (blue dashed-dotted lines) are also displayed.
      }
    \label{fig:xmm_final}
   \end{suppfigure*}

 \begin{suppfigure*}
   \centering
   \includegraphics[width=0.99\columnwidth]{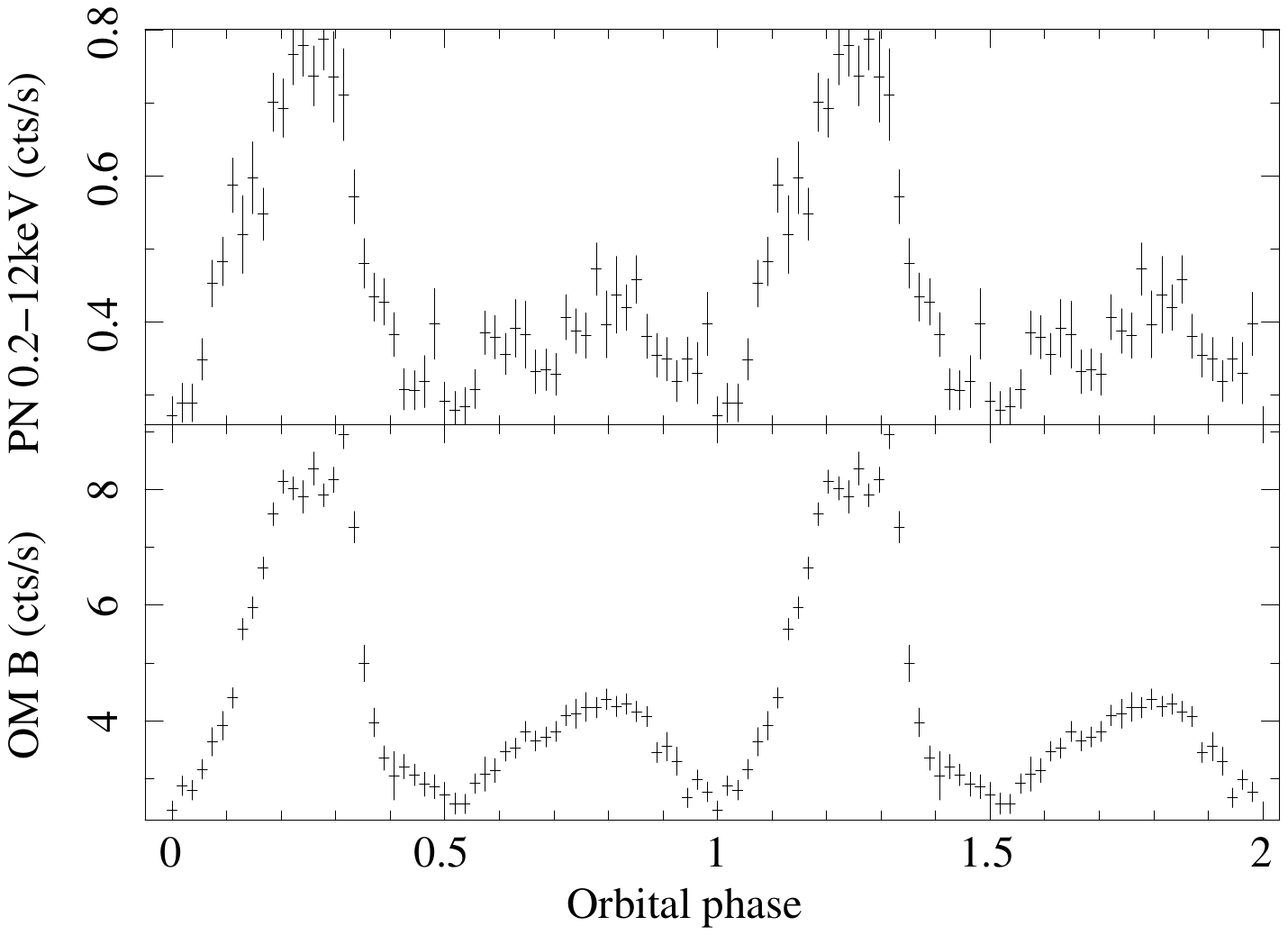}
   \includegraphics[width=0.99\columnwidth]{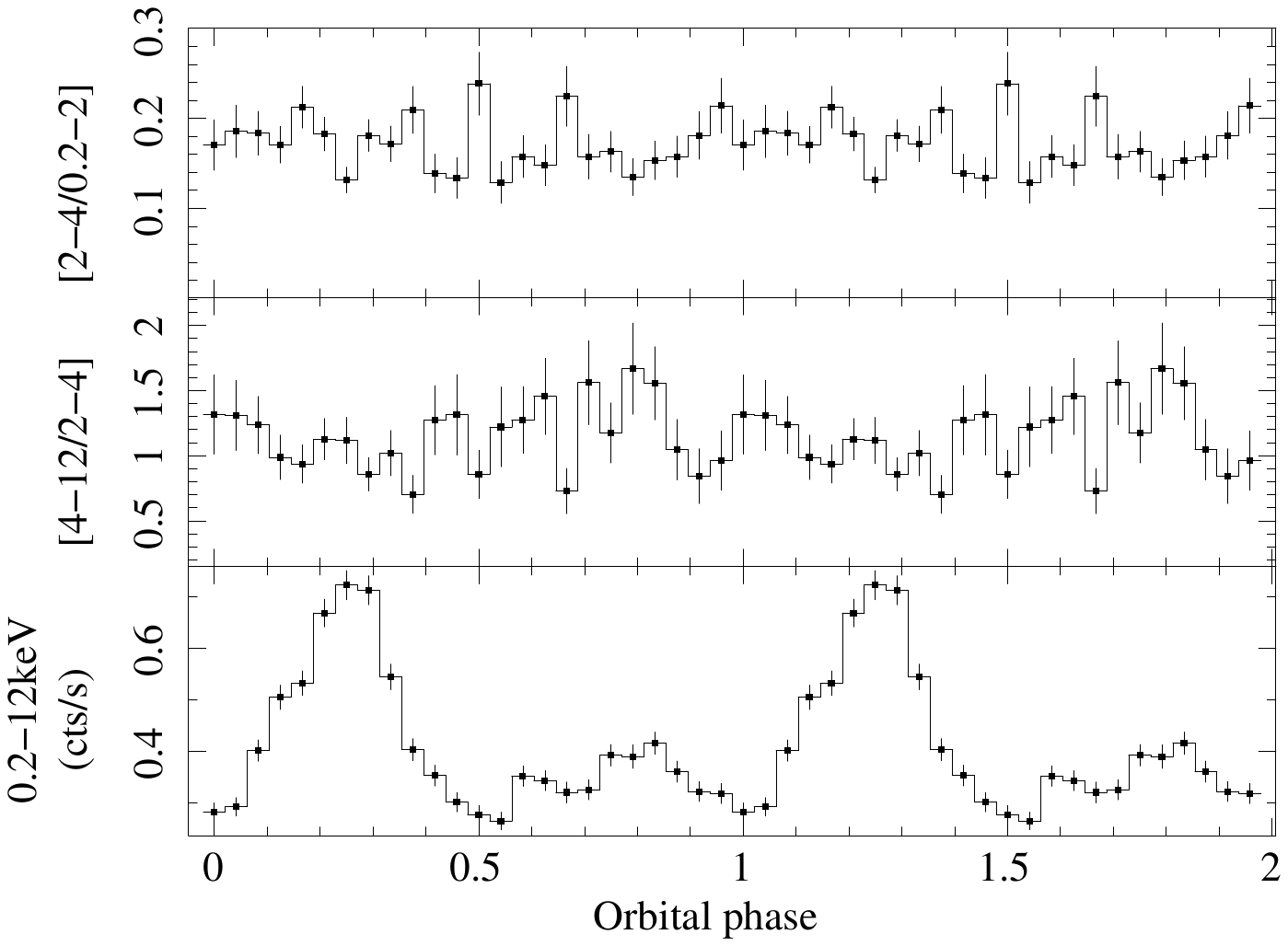}
      \caption{
      \textbf{Folded X-ray and optical lightcurves of RXJ0528+2838.} {\it Left: } The X-ray and optical light curves of J0528+3828 folded at the 80\,m orbital period sampled with 54 bins. {\it Right:} Folded hardness ratios at the orbital period in selected 0.2-2\,keV, 2-4\,keV and 4-12\,keV bands evaluated in 24 bins. The light curve in the total 0.2-12\,keV range is displayed at the bottom for comparison. All error bars display the $1\sigma$ uncertainty.}
    \label{fig:folded_xmm}
   \end{suppfigure*}

\begin{suppfigure*}
   \centering
   \includegraphics[width=0.99\columnwidth]{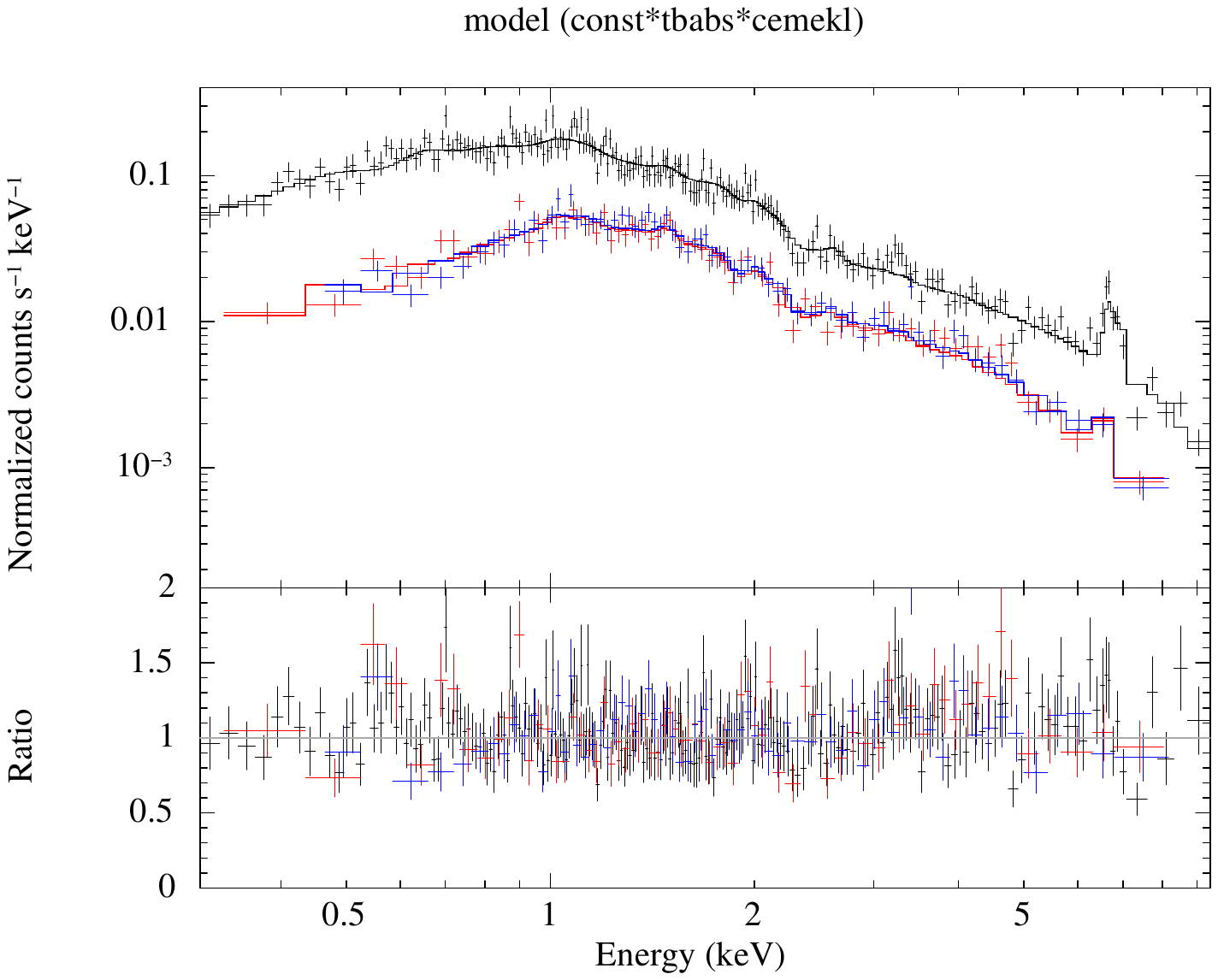}
      \caption{\textbf{X-ray spectrum of RXJ0528+2838.} The X-ray spectra from the EPIC-pn (black) and MOS1 (red) and MOS2 (green) instruments fitted with an absorbed multi-temperature plasma (upper panel). The ratio between observed and model is reported in the lower panel. All error bars display the $1\sigma$ uncertainty.}
    \label{fig:xmm_spec}
   \end{suppfigure*}

   \begin{suppfigure*}
   \centering
   \includegraphics[width=0.99\columnwidth]{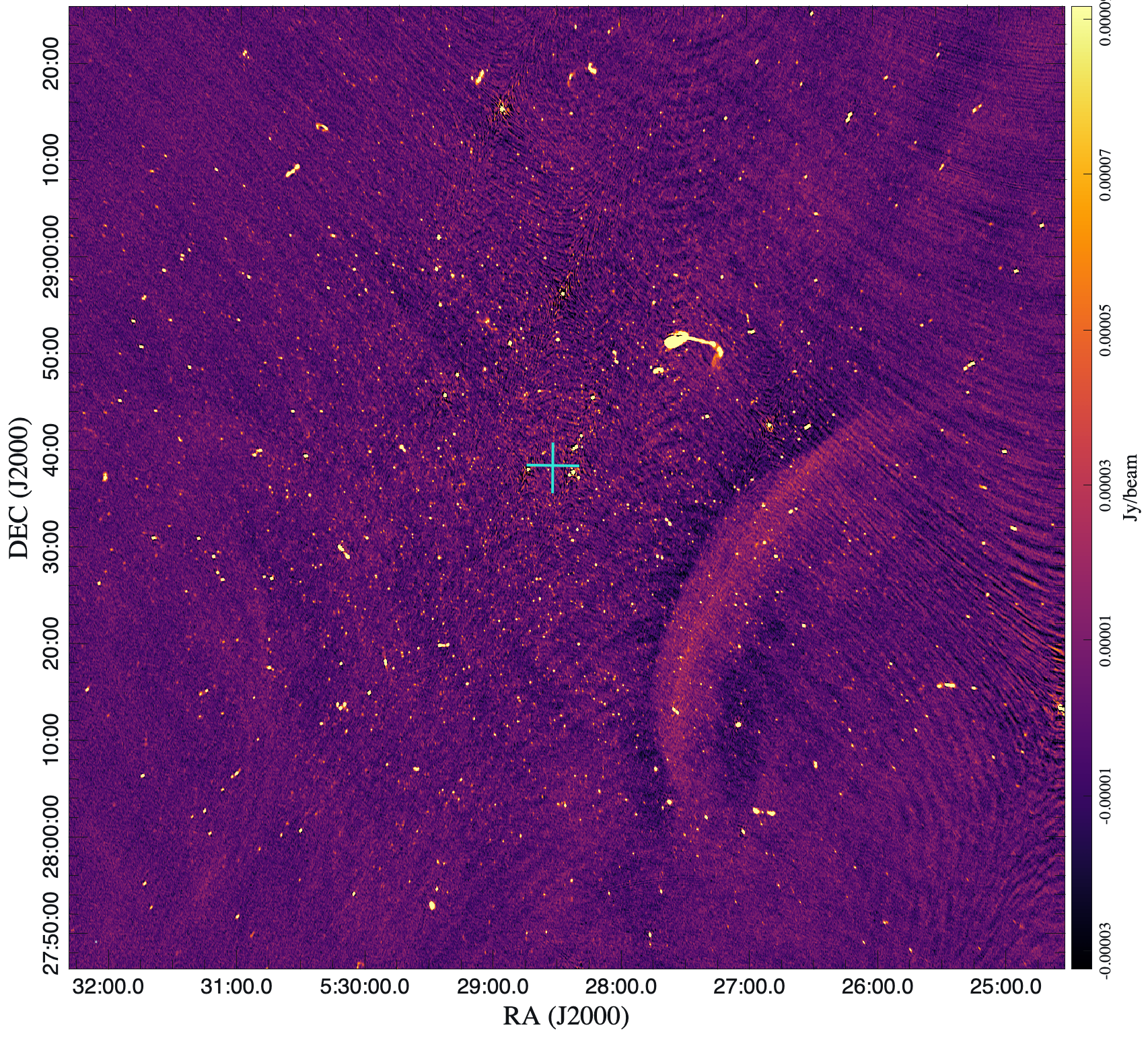}
      \caption{\textbf{The field of RXJ0528+2838 as seen by the MeerKAT radio telescope.} The diamond at the geometrical centre of the $1.5$ degrees$^2$ field corresponds to the nominal position of RXJ0528+2838. }
         \label{fig:MeerKAT_field}
   \end{suppfigure*}


\begin{thebibliography}{10}
\expandafter\ifx\csname url\endcsname\relax
  \def\url#1{\burl{#1}}\fi
\expandafter\ifx\csname urlprefix\endcsname\relax\def\urlprefix{URL }\fi
\providecommand{\bibinfo}[2]{#2}
\providecommand{\eprint}[2][]{\url{#2}}
\providecommand{\doi}[1]{\url{https://doi.org/#1}}

\bibitem{2015AstBu..70..460G}
\bibinfo{author}{{Gabdeev}, M.~M.}
\newblock \bibinfo{title}{{Photometric monitoring of polar candidates}}.
\newblock \emph{\bibinfo{journal}{Astrophysical Bulletin}}
  \textbf{\bibinfo{volume}{70}}, \bibinfo{pages}{460--465}
  (\bibinfo{year}{2015}).

\bibitem{2021AJ....161..147B}
\bibinfo{author}{{Bailer-Jones}, C.~A.~L.}, \bibinfo{author}{{Rybizki}, J.},
  \bibinfo{author}{{Fouesneau}, M.}, \bibinfo{author}{{Demleitner}, M.} \&
  \bibinfo{author}{{Andrae}, R.}
\newblock \bibinfo{title}{{Estimating Distances from Parallaxes. V. Geometric
  and Photogeometric Distances to 1.47 Billion Stars in Gaia Early Data Release
  3}}.
\newblock \emph{\bibinfo{journal}{\aj}} \textbf{\bibinfo{volume}{161}},
  \bibinfo{pages}{147} (\bibinfo{year}{2021}).

\bibitem{2016AstBu..71...95B}
\bibinfo{author}{{Borisov}, N.~V.}, \bibinfo{author}{{Gabdeev}, M.~M.} \&
  \bibinfo{author}{{Afanasiev}, V.~L.}
\newblock \bibinfo{title}{{Photopolarimetric observations of the sample of
  polar candidates}}.
\newblock \emph{\bibinfo{journal}{Astrophysical Bulletin}}
  \textbf{\bibinfo{volume}{71}}, \bibinfo{pages}{95--100}
  (\bibinfo{year}{2016}).

\bibitem{2021ApJS..257...65S}
\bibinfo{author}{{Sun}, Y.} \emph{et~al.}
\newblock \bibinfo{title}{{A Catalog of 323 Cataclysmic Variables from LAMOST
  DR6}}.
\newblock \emph{\bibinfo{journal}{\apjs}} \textbf{\bibinfo{volume}{257}},
  \bibinfo{pages}{65} (\bibinfo{year}{2021}).

\bibitem{2005MNRAS.362..753D}
\bibinfo{author}{{Drew}, J.~E.} \emph{et~al.}
\newblock \bibinfo{title}{{The INT Photometric H{\ensuremath{\alpha}} Survey of
  the Northern Galactic Plane (IPHAS)}}.
\newblock \emph{\bibinfo{journal}{\mnras}} \textbf{\bibinfo{volume}{362}},
  \bibinfo{pages}{753--776} (\bibinfo{year}{2005}).

\bibitem{2021A&A...655A..49G}
\bibinfo{author}{{Greimel}, R.} \emph{et~al.}
\newblock \bibinfo{title}{{High-resolution H{\ensuremath{\alpha}} imaging of
  the northern Galactic plane and the IGAPS image database}}.
\newblock \emph{\bibinfo{journal}{\aap}} \textbf{\bibinfo{volume}{655}},
  \bibinfo{pages}{A49} (\bibinfo{year}{2021}).

\bibitem{2021A&A...649A...1G}
\bibinfo{author}{{Gaia Collaboration}} \emph{et~al.}
\newblock \bibinfo{title}{{Gaia Early Data Release 3. Summary of the contents
  and survey properties}}.
\newblock \emph{\bibinfo{journal}{\aap}} \textbf{\bibinfo{volume}{649}},
  \bibinfo{pages}{A1} (\bibinfo{year}{2021}).

\bibitem{2020MNRAS.494.3799P}
\bibinfo{author}{{Pala}, A.~F.} \emph{et~al.}
\newblock \bibinfo{title}{{A Volume-limited Sample of Cataclysmic Variables
  from Gaia DR2: Space Density and Population Properties}}.
\newblock \emph{\bibinfo{journal}{\mnras}} \textbf{\bibinfo{volume}{494}},
  \bibinfo{pages}{3799--3827} (\bibinfo{year}{2020}).

\bibitem{2025PASP..137a4201R}
\bibinfo{author}{{Rodriguez}, A.~C.} \emph{et~al.}
\newblock \bibinfo{title}{{Cataclysmic Variables and AM CVn Binaries in
  SRG/eROSITA + Gaia: Volume Limited Samples, X-Ray Luminosity Functions, and
  Space Densities}}.
\newblock \emph{\bibinfo{journal}{\pasp}} \textbf{\bibinfo{volume}{137}},
  \bibinfo{pages}{014201} (\bibinfo{year}{2025}).

\bibitem{1995MNRAS.276..353S}
\bibinfo{author}{{Slavin}, A.~J.}, \bibinfo{author}{{O'Brien}, T.~J.} \&
  \bibinfo{author}{{Dunlop}, J.~S.}
\newblock \bibinfo{title}{{A deep optical imaging study of the nebular remnants
  of classical novae}}.
\newblock \emph{\bibinfo{journal}{\mnras}} \textbf{\bibinfo{volume}{276}},
  \bibinfo{pages}{353--371} (\bibinfo{year}{1995}).

\bibitem{1987A&A...181..373K}
\bibinfo{author}{{Krautter}, J.}, \bibinfo{author}{{Klaas}, U.} \&
  \bibinfo{author}{{Radons}, G.}
\newblock \bibinfo{title}{{On the Nature of 623+71: A Cataclysmic Binary
  Surrounded by a Bow Shock-like Emission Nebula}}.
\newblock \emph{\bibinfo{journal}{\aap}} \textbf{\bibinfo{volume}{181}},
  \bibinfo{pages}{373--377} (\bibinfo{year}{1987}).

\bibitem{1992ApJ...393..217H}
\bibinfo{author}{{Hollis}, J.~M.}, \bibinfo{author}{{Oliversen}, R.~J.},
  \bibinfo{author}{{Wagner}, R.~M.} \& \bibinfo{author}{{Feibelman}, W.~A.}
\newblock \bibinfo{title}{{The 0623+71 Bow Shock Nebula}}.
\newblock \emph{\bibinfo{journal}{\apj}} \textbf{\bibinfo{volume}{393}},
  \bibinfo{pages}{217} (\bibinfo{year}{1992}).

\bibitem{2001A&A...376.1031G}
\bibinfo{author}{{Greiner}, J.} \emph{et~al.}
\newblock \bibinfo{title}{{BZ Camelopardalis during its 1999/2000 optical low
  state}}.
\newblock \emph{\bibinfo{journal}{\aap}} \textbf{\bibinfo{volume}{376}},
  \bibinfo{pages}{1031--1038} (\bibinfo{year}{2001}).

\bibitem{2008PhDT.......109F}
\bibinfo{author}{{Frew}, D.~J.}
\newblock \emph{\bibinfo{title}{{Planetary Nebulae in the Solar Neighbourhood:
  Statistics, Distance Scale and Luminosity Function}}}.
\newblock Ph.D. thesis, \bibinfo{school}{Macquarie University, Department of
  Physics and Astronomy} (\bibinfo{year}{2008}).

\bibitem{2018PASP..130i4201B}
\bibinfo{author}{{Bond}, H.~E.} \& \bibinfo{author}{{Miszalski}, B.}
\newblock \bibinfo{title}{{Spectroscopy of V341 Arae: A Nearby Nova-like
  Variable Inside a Bow Shock Nebula}}.
\newblock \emph{\bibinfo{journal}{\pasp}} \textbf{\bibinfo{volume}{130}},
  \bibinfo{pages}{094201} (\bibinfo{year}{2018}).

\bibitem{2019MNRAS.486.2631H}
\bibinfo{author}{{Hern{\'a}ndez Santisteban}, J.~V.} \emph{et~al.}
\newblock \bibinfo{title}{{From outburst to quiescence: spectroscopic evolution
  of V1838 Aql imbedded in a bow-shock nebula}}.
\newblock \emph{\bibinfo{journal}{\mnras}} \textbf{\bibinfo{volume}{486}},
  \bibinfo{pages}{2631--2642} (\bibinfo{year}{2019}).

\bibitem{2021MNRAS.501.1951C}
\bibinfo{author}{{Castro Segura}, N.} \emph{et~al.}
\newblock \bibinfo{title}{{Bow shocks, nova shells, disc winds and tilted
  discs: the nova-like V341 Ara has it all}}.
\newblock \emph{\bibinfo{journal}{\mnras}} \textbf{\bibinfo{volume}{501}},
  \bibinfo{pages}{1951--1969} (\bibinfo{year}{2021}).

\bibitem{2024AJ....168..249B}
\bibinfo{author}{{Bond}, H.~E.} \emph{et~al.}
\newblock \bibinfo{title}{{Discovery of a Bow-shock Nebula Around the Z
  Cam-type Cataclysmic Variable SY Cancri}}.
\newblock \emph{\bibinfo{journal}{\aj}} \textbf{\bibinfo{volume}{168}},
  \bibinfo{pages}{249} (\bibinfo{year}{2024}).

\bibitem{brownsberger14}
\bibinfo{author}{{Brownsberger}, S.} \& \bibinfo{author}{{Romani}, R.~W.}
\newblock \bibinfo{title}{{A Survey for H{\ensuremath{\alpha}} Pulsar Bow
  Shocks}}.
\newblock \emph{\bibinfo{journal}{\apj}} \textbf{\bibinfo{volume}{784}},
  \bibinfo{pages}{154} (\bibinfo{year}{2014}).

\bibitem{bhalerao19}
\bibinfo{author}{{Bhalerao}, J.}, \bibinfo{author}{{Park}, S.},
  \bibinfo{author}{{Schenck}, A.}, \bibinfo{author}{{Post}, S.} \&
  \bibinfo{author}{{Hughes}, J.~P.}
\newblock \bibinfo{title}{{Detailed X-Ray Mapping of the Shocked Ejecta and
  Circumstellar Medium in the Galactic Core-collapse Supernova Remnant
  G292.0+1.8}}.
\newblock \emph{\bibinfo{journal}{\apj}} \textbf{\bibinfo{volume}{872}},
  \bibinfo{pages}{31} (\bibinfo{year}{2019}).

\bibitem{2016A&A...595A..64H}
\bibinfo{author}{{Harvey}, E.}, \bibinfo{author}{{Redman}, M.~P.},
  \bibinfo{author}{{Boumis}, P.} \& \bibinfo{author}{{Akras}, S.}
\newblock \bibinfo{title}{{Modelling the structure and kinematics of the
  Firework nebula: The nature of the GK Persei nova shell and its jet-like
  feature}}.
\newblock \emph{\bibinfo{journal}{\aap}} \textbf{\bibinfo{volume}{595}},
  \bibinfo{pages}{A64} (\bibinfo{year}{2016}).

\bibitem{2024ApJ...972L..14I}
\bibinfo{author}{{I{\l}kiewicz}, K.}, \bibinfo{author}{{Miko{\l}ajewska}, J.},
  \bibinfo{author}{{Shara}, M.~M.}, \bibinfo{author}{{Faherty}, J.~K.} \&
  \bibinfo{author}{{Scaringi}, S.}
\newblock \bibinfo{title}{{Ancient Nova Shells of RX Pup Indicate Evolution of
  Mass Transfer Rate}}.
\newblock \emph{\bibinfo{journal}{\apjl}} \textbf{\bibinfo{volume}{972}},
  \bibinfo{pages}{L14} (\bibinfo{year}{2024}).

\bibitem{2024A&A...681A.106C}
\bibinfo{author}{{Celed{\'o}n}, L.}, \bibinfo{author}{{Schmidtobreick}, L.},
  \bibinfo{author}{{Tappert}, C.} \& \bibinfo{author}{{Selman}, F.}
\newblock \bibinfo{title}{{Unveiling the 3D structure of nova shells with MUSE:
  The case of RR Pic}}.
\newblock \emph{\bibinfo{journal}{\aap}} \textbf{\bibinfo{volume}{681}},
  \bibinfo{pages}{A106} (\bibinfo{year}{2024}).

\bibitem{celedon25}
\bibinfo{author}{{Celed{\'o}n}, L.}, \bibinfo{author}{{Tappert}, C.},
  \bibinfo{author}{{Schmidtobreick}, L.} \& \bibinfo{author}{{Selman}, F.~J.}
\newblock \bibinfo{title}{{MUSE observations of V1425 Aql reveal an arc-shaped
  nova shell}}.
\newblock \emph{\bibinfo{journal}{\aap}} \textbf{\bibinfo{volume}{694}},
  \bibinfo{pages}{A238} (\bibinfo{year}{2025}).

\bibitem{2007MNRAS.380..175V}
\bibinfo{author}{{Vaytet}, N.~M.~H.}, \bibinfo{author}{{O'Brien}, T.~J.} \&
  \bibinfo{author}{{Rushton}, A.~P.}
\newblock \bibinfo{title}{{Evidence for ablated flows in the shell of the nova
  DQ Herculis}}.
\newblock \emph{\bibinfo{journal}{\mnras}} \textbf{\bibinfo{volume}{380}},
  \bibinfo{pages}{175--180} (\bibinfo{year}{2007}).

\bibitem{2022MNRAS.517.2567S}
\bibinfo{author}{{Santamar{\'\i}a}, E.}, \bibinfo{author}{{Guerrero}, M.~A.},
  \bibinfo{author}{{Toal{\'a}}, J.~A.}, \bibinfo{author}{{Ramos-Larios}, G.} \&
  \bibinfo{author}{{Sabin}, L.}
\newblock \bibinfo{title}{{QU Vul: An integral field spectroscopy case study of
  a nova shell}}.
\newblock \emph{\bibinfo{journal}{\mnras}} \textbf{\bibinfo{volume}{517}},
  \bibinfo{pages}{2567--2576} (\bibinfo{year}{2022}).

\bibitem{1977ApJ...218..377W}
\bibinfo{author}{{Weaver}, R.}, \bibinfo{author}{{McCray}, R.},
  \bibinfo{author}{{Castor}, J.}, \bibinfo{author}{{Shapiro}, P.} \&
  \bibinfo{author}{{Moore}, R.}
\newblock \bibinfo{title}{{Interstellar bubbles. II. Structure and evolution.}}
\newblock \emph{\bibinfo{journal}{\apj}} \textbf{\bibinfo{volume}{218}},
  \bibinfo{pages}{377--395} (\bibinfo{year}{1977}).

\bibitem{2017ApJ...835...29Y}
\bibinfo{author}{{Yao}, J.~M.}, \bibinfo{author}{{Manchester}, R.~N.} \&
  \bibinfo{author}{{Wang}, N.}
\newblock \bibinfo{title}{{A New Electron-density Model for Estimation of
  Pulsar and FRB Distances}}.
\newblock \emph{\bibinfo{journal}{\apj}} \textbf{\bibinfo{volume}{835}},
  \bibinfo{pages}{29} (\bibinfo{year}{2017}).

\bibitem{2013ApJ...773L..11F}
\bibinfo{author}{{Foster}, T.}, \bibinfo{author}{{Kothes}, R.} \&
  \bibinfo{author}{{Brown}, J.~C.}
\newblock \bibinfo{title}{{A Relation between the Warm Neutral and Ionized
  Media Observed in the Canadian Galactic Plane Survey}}.
\newblock \emph{\bibinfo{journal}{\apjl}} \textbf{\bibinfo{volume}{773}},
  \bibinfo{pages}{L11} (\bibinfo{year}{2013}).

\bibitem{2013ApJ...764...25J}
\bibinfo{author}{{Jenkins}, E.~B.}
\newblock \bibinfo{title}{{The Fractional Ionization of the Warm Neutral
  Interstellar Medium}}.
\newblock \emph{\bibinfo{journal}{\apj}} \textbf{\bibinfo{volume}{764}},
  \bibinfo{pages}{25} (\bibinfo{year}{2013}).

\bibitem{2025AJ....170...78B}
\bibinfo{author}{{Bond}, H.~E.} \emph{et~al.}
\newblock \bibinfo{title}{{Two More Bow Shocks and Off-center
  H{\ensuremath{\alpha}} Nebulae Associated with Nova-like Cataclysmic
  Variables}}.
\newblock \emph{\bibinfo{journal}{\aj}} \textbf{\bibinfo{volume}{170}},
  \bibinfo{pages}{78} (\bibinfo{year}{2025}).

\bibitem{1990SSRv...54..195C}
\bibinfo{author}{{Cropper}, M.}
\newblock \bibinfo{title}{{The Polars}}.
\newblock \emph{\bibinfo{journal}{\ssr}} \textbf{\bibinfo{volume}{54}},
  \bibinfo{pages}{195--295} (\bibinfo{year}{1990}).

\bibitem{2022MNRAS.510.6110P}
\bibinfo{author}{{Pala}, A.~F.} \emph{et~al.}
\newblock \bibinfo{title}{{Constraining the evolution of cataclysmic variables
  via the masses and accretion rates of their underlying white dwarfs}}.
\newblock \emph{\bibinfo{journal}{\mnras}} \textbf{\bibinfo{volume}{510}},
  \bibinfo{pages}{6110--6132} (\bibinfo{year}{2022}).

\bibitem{2011ApJS..194...28K}
\bibinfo{author}{{Knigge}, C.}, \bibinfo{author}{{Baraffe}, I.} \&
  \bibinfo{author}{{Patterson}, J.}
\newblock \bibinfo{title}{{The Evolution of Cataclysmic Variables as Revealed
  by Their Donor Stars}}.
\newblock \emph{\bibinfo{journal}{\apjs}} \textbf{\bibinfo{volume}{194}},
  \bibinfo{pages}{28} (\bibinfo{year}{2011}).

\bibitem{mason24}
\bibinfo{author}{{Mason}, P.~A.} \emph{et~al.}
\newblock \bibinfo{title}{{TESS Photometry of AM Her and AR UMa: Binary
  Parameters, Cyclotron Emission Modeling, and Mass Transfer Duty Cycles}}.
\newblock \emph{\bibinfo{journal}{\apj}} \textbf{\bibinfo{volume}{965}},
  \bibinfo{pages}{96} (\bibinfo{year}{2024}).

\bibitem{pagnotta16}
\bibinfo{author}{{Pagnotta}, A.} \& \bibinfo{author}{{Zurek}, D.}
\newblock \bibinfo{title}{{Non-detection of nova shells around asynchronous
  polars}}.
\newblock \emph{\bibinfo{journal}{\mnras}} \textbf{\bibinfo{volume}{458}},
  \bibinfo{pages}{1833--1838} (\bibinfo{year}{2016}).

\bibitem{ridder23}
\bibinfo{author}{{Ridder}, M.~E.}, \bibinfo{author}{{Heinke}, C.~O.},
  \bibinfo{author}{{Sivakoff}, G.~R.} \& \bibinfo{author}{{Hughes}, A.~K.}
\newblock \bibinfo{title}{{Radio detections of two unusual cataclysmic
  variables in the VLA Sky Survey}}.
\newblock \emph{\bibinfo{journal}{\mnras}} \textbf{\bibinfo{volume}{519}},
  \bibinfo{pages}{5922--5930} (\bibinfo{year}{2023}).

\bibitem{marsh16}
\bibinfo{author}{{Marsh}, T.~R.} \emph{et~al.}
\newblock \bibinfo{title}{{A radio-pulsing white dwarf binary star}}.
\newblock \emph{\bibinfo{journal}{\nat}} \textbf{\bibinfo{volume}{537}},
  \bibinfo{pages}{374--377} (\bibinfo{year}{2016}).

\bibitem{deRuiter25}
\bibinfo{author}{{de Ruiter}, I.} \emph{et~al.}
\newblock \bibinfo{title}{{Sporadic radio pulses from a white dwarf binary at
  the orbital period}}.
\newblock \emph{\bibinfo{journal}{Nature Astronomy}}  (\bibinfo{year}{2025}).

\bibitem{schreiber21}
\bibinfo{author}{{Schreiber}, M.~R.}, \bibinfo{author}{{Belloni}, D.},
  \bibinfo{author}{{G{\"a}nsicke}, B.~T.}, \bibinfo{author}{{Parsons}, S.~G.}
  \& \bibinfo{author}{{Zorotovic}, M.}
\newblock \bibinfo{title}{{The origin and evolution of magnetic white dwarfs in
  close binary stars}}.
\newblock \emph{\bibinfo{journal}{Nature Astronomy}}
  \textbf{\bibinfo{volume}{5}}, \bibinfo{pages}{648--654}
  (\bibinfo{year}{2021}).

\bibitem{rodriguez25}
\bibinfo{author}{{Rodriguez}, A.~C.}
\newblock \bibinfo{title}{{Spectroscopic detection of a 2.9-hour orbit in a
  long-period radio transient}}.
\newblock \emph{\bibinfo{journal}{\aap}} \textbf{\bibinfo{volume}{695}},
  \bibinfo{pages}{L8} (\bibinfo{year}{2025}).

\bibitem{castroSegura25}
\bibinfo{author}{{Castro Segura}, N.} \emph{et~al.}
\newblock \bibinfo{title}{{A sibling of AR Scorpii: SDSS J230641.47+244055.8
  and the observational blueprint of white dwarf pulsars}}.
\newblock \emph{\bibinfo{journal}{\mnras}} \textbf{\bibinfo{volume}{543}},
  \bibinfo{pages}{2116--2129} (\bibinfo{year}{2025}).

\bibitem{ferrario15}
\bibinfo{author}{{Ferrario}, L.}, \bibinfo{author}{{de Martino}, D.} \&
  \bibinfo{author}{{G{\"a}nsicke}, B.~T.}
\newblock \bibinfo{title}{{Magnetic White Dwarfs}}.
\newblock \emph{\bibinfo{journal}{\ssr}} \textbf{\bibinfo{volume}{191}},
  \bibinfo{pages}{111--169} (\bibinfo{year}{2015}).





\bibitem{2015JATIS...1a4003R}
\bibinfo{author}{{Ricker}, G.~R.} \emph{et~al.}
\newblock \bibinfo{title}{{Transiting Exoplanet Survey Satellite (TESS)}}.
\newblock \emph{\bibinfo{journal}{Journal of Astronomical Telescopes,
  Instruments, and Systems}} \textbf{\bibinfo{volume}{1}},
  \bibinfo{pages}{014003} (\bibinfo{year}{2015}).

\bibitem{1999DSSN...13...28M}
\bibinfo{author}{{Montgomery}, M.~H.} \& \bibinfo{author}{{O'Donoghue}, D.}
\newblock \bibinfo{title}{{A derivation of the errors for least squares fitting
  to time series data}}.
\newblock \emph{\bibinfo{journal}{Delta Scuti Star Newsletter}}
  \textbf{\bibinfo{volume}{13}}, \bibinfo{pages}{28} (\bibinfo{year}{1999}).

\bibitem{2022MNRAS.516.5209I}
\bibinfo{author}{{I{\l}kiewicz}, K.}, \bibinfo{author}{{Scaringi}, S.},
  \bibinfo{author}{{Littlefield}, C.} \& \bibinfo{author}{{Mason}, P.~A.}
\newblock \bibinfo{title}{{Locating the flickering source in polars}}.
\newblock \emph{\bibinfo{journal}{\mnras}} \textbf{\bibinfo{volume}{516}},
  \bibinfo{pages}{5209--5215} (\bibinfo{year}{2022}).

\bibitem{2019MNRAS.489.1044B}
\bibinfo{author}{{Bernardini}, F.}, \bibinfo{author}{{de Martino}, D.},
  \bibinfo{author}{{Mukai}, K.}, \bibinfo{author}{{Falanga}, M.} \&
  \bibinfo{author}{{Masetti}, N.}
\newblock \bibinfo{title}{{2PBC J0658.0-1746: a hard X-ray eclipsing polar in
  the orbital period gap}}.
\newblock \emph{\bibinfo{journal}{\mnras}} \textbf{\bibinfo{volume}{489}},
  \bibinfo{pages}{1044--1053} (\bibinfo{year}{2019}).

\bibitem{1994PASP..106..209P}
\bibinfo{author}{{Patterson}, J.}
\newblock \bibinfo{title}{{The DQ Herculis Stars}}.
\newblock \emph{\bibinfo{journal}{\pasp}} \textbf{\bibinfo{volume}{106}},
  \bibinfo{pages}{209} (\bibinfo{year}{1994}).

\bibitem{2014ApJ...788...48S}
\bibinfo{author}{{Shappee}, B.~J.} \emph{et~al.}
\newblock \bibinfo{title}{{The Man behind the Curtain: X-Rays Drive the UV
  through NIR Variability in the 2013 Active Galactic Nucleus Outburst in NGC
  2617}}.
\newblock \emph{\bibinfo{journal}{\apj}} \textbf{\bibinfo{volume}{788}},
  \bibinfo{pages}{48} (\bibinfo{year}{2014}).

\bibitem{2017PASP..129j4502K}
\bibinfo{author}{{Kochanek}, C.~S.} \emph{et~al.}
\newblock \bibinfo{title}{{The All-Sky Automated Survey for Supernovae
  (ASAS-SN) Light Curve Server v1.0}}.
\newblock \emph{\bibinfo{journal}{\pasp}} \textbf{\bibinfo{volume}{129}},
  \bibinfo{pages}{104502} (\bibinfo{year}{2017}).

\bibitem{2018PASP..130f4505T}
\bibinfo{author}{{Tonry}, J.~L.} \emph{et~al.}
\newblock \bibinfo{title}{{ATLAS: A High-cadence All-sky Survey System}}.
\newblock \emph{\bibinfo{journal}{\pasp}} \textbf{\bibinfo{volume}{130}},
  \bibinfo{pages}{064505} (\bibinfo{year}{2018}).

\bibitem{2018AJ....156..241H}
\bibinfo{author}{{Heinze}, A.~N.} \emph{et~al.}
\newblock \bibinfo{title}{{A First Catalog of Variable Stars Measured by the
  Asteroid Terrestrial-impact Last Alert System (ATLAS)}}.
\newblock \emph{\bibinfo{journal}{\aj}} \textbf{\bibinfo{volume}{156}},
  \bibinfo{pages}{241} (\bibinfo{year}{2018}).

\bibitem{2010AJ....140.1868W}
\bibinfo{author}{{Wright}, E.~L.} \emph{et~al.}
\newblock \bibinfo{title}{{The Wide-field Infrared Survey Explorer (WISE):
  Mission Description and Initial On-orbit Performance}}.
\newblock \emph{\bibinfo{journal}{\aj}} \textbf{\bibinfo{volume}{140}},
  \bibinfo{pages}{1868--1881} (\bibinfo{year}{2010}).

\bibitem{2011ApJ...731...53M}
\bibinfo{author}{{Mainzer}, A.} \emph{et~al.}
\newblock \bibinfo{title}{{Preliminary Results from NEOWISE: An Enhancement to
  the Wide-field Infrared Survey Explorer for Solar System Science}}.
\newblock \emph{\bibinfo{journal}{\apj}} \textbf{\bibinfo{volume}{731}},
  \bibinfo{pages}{53} (\bibinfo{year}{2011}).

\bibitem{2020MNRAS.493.2271H}
\bibinfo{author}{{Hwang}, H.-C.} \& \bibinfo{author}{{Zakamska}, N.~L.}
\newblock \bibinfo{title}{{Lifetime of short-period binaries measured from
  their Galactic kinematics}}.
\newblock \emph{\bibinfo{journal}{\mnras}} \textbf{\bibinfo{volume}{493}},
  \bibinfo{pages}{2271--2286} (\bibinfo{year}{2020}).

\bibitem{2008A&A...481..433W}
\bibinfo{author}{{Wu}, K.} \& \bibinfo{author}{{Kiss}, L.~L.}
\newblock \bibinfo{title}{{High and low states of the system AM Herculis}}.
\newblock \emph{\bibinfo{journal}{\aap}} \textbf{\bibinfo{volume}{481}},
  \bibinfo{pages}{433--439} (\bibinfo{year}{2008}).

\bibitem{2010SPIE.7735E..08B}
\bibinfo{author}{{Bacon}, R.} \emph{et~al.}
\newblock \bibinfo{editor}{{McLean}, I.~S.}, \bibinfo{editor}{{Ramsay}, S.~K.}
  \& \bibinfo{editor}{{Takami}, H.} (eds) \emph{\bibinfo{title}{{The MUSE
  second-generation VLT instrument}}}.
\newblock (eds \bibinfo{editor}{{McLean}, I.~S.}, \bibinfo{editor}{{Ramsay},
  S.~K.} \& \bibinfo{editor}{{Takami}, H.})
  \emph{\bibinfo{booktitle}{Ground-based and Airborne Instrumentation for
  Astronomy III}}, Vol. \bibinfo{volume}{7735} of
  \emph{\bibinfo{series}{Society of Photo-Optical Instrumentation Engineers
  (SPIE) Conference Series}}, \bibinfo{pages}{773508} (\bibinfo{year}{2010}).
\newblock
  \bibinfo{eprint}{{\href{https://arxiv.org/abs/2211.16795}{{arXiv:2211.16795}}}}.

\bibitem{2020A&A...641A..28W}
\bibinfo{author}{{Weilbacher}, P.~M.} \emph{et~al.}
\newblock \bibinfo{title}{{The data processing pipeline for the MUSE
  instrument}}.
\newblock \emph{\bibinfo{journal}{\aap}} \textbf{\bibinfo{volume}{641}},
  \bibinfo{pages}{A28} (\bibinfo{year}{2020}).

\bibitem{2022AJ....163..291G}
\bibinfo{author}{{Ginsburg}, A.} \emph{et~al.}
\newblock \bibinfo{title}{{Pyspeckit: A Spectroscopic Analysis and Plotting
  Package}}.
\newblock \emph{\bibinfo{journal}{\aj}} \textbf{\bibinfo{volume}{163}},
  \bibinfo{pages}{291} (\bibinfo{year}{2022}).

\bibitem{2011ascl.soft09001G}
\bibinfo{author}{Ginsburg, A.} \& \bibinfo{author}{Mirocha, J.}
\newblock \bibinfo{title}{Pyspeckit: Python spectroscopic toolkit}.
\newblock \bibinfo{howpublished}{Astrophysics Source Code Library, record
  ascl:1109.001} (\bibinfo{year}{2011}).

\bibitem{1999ASPC..157..127B}
\bibinfo{author}{{Burwitz}, V.}, \bibinfo{author}{{Reinsch}, K.},
  \bibinfo{author}{{Beuermann}, K.} \& \bibinfo{author}{{Thomas}, H.-C.}
\newblock \bibinfo{editor}{{Hellier}, C.} \& \bibinfo{editor}{{Mukai}, K.}
  (eds) \emph{\bibinfo{title}{{RX J0501.7-0359: a new ROSAT discovered
  eclipsing polar in the period gap}}}.
\newblock (eds \bibinfo{editor}{{Hellier}, C.} \& \bibinfo{editor}{{Mukai},
  K.}) \emph{\bibinfo{booktitle}{Annapolis Workshop on Magnetic Cataclysmic
  Variables}}, Vol. \bibinfo{volume}{157} of
  \emph{\bibinfo{series}{Astronomical Society of the Pacific Conference
  Series}}, \bibinfo{pages}{127} (\bibinfo{year}{1999}).
\newblock
  \bibinfo{eprint}{{\href{https://arxiv.org/abs/astro-ph/9810437}{{arXiv:astro-ph/9810437}}}}.

\bibitem{2019ASPC..518..100G}
\bibinfo{author}{{Gabdeev}, M.~M.}, \bibinfo{author}{{Borisov}, N.~V.},
  \bibinfo{author}{{Shimansky}, V.~V.}, \bibinfo{author}{{Kolbin}, A.~I.} \&
  \bibinfo{author}{{Nikolaeva}, E.~A.}
\newblock \bibinfo{editor}{{Kudryavtsev}, D.~O.}, \bibinfo{editor}{{Romanyuk},
  I.~I.} \& \bibinfo{editor}{{Yakunin}, I.~A.} (eds) \emph{\bibinfo{title}{{A
  Spectroscopic Study of a New Magnetic Cataclysmic Variable IPHAS
  J052832.69+283837.6}}}.
\newblock (eds \bibinfo{editor}{{Kudryavtsev}, D.~O.},
  \bibinfo{editor}{{Romanyuk}, I.~I.} \& \bibinfo{editor}{{Yakunin}, I.~A.})
  \emph{\bibinfo{booktitle}{Physics of Magnetic Stars}}, Vol.
  \bibinfo{volume}{518} of \emph{\bibinfo{series}{Astronomical Society of the
  Pacific Conference Series}}, \bibinfo{pages}{100} (\bibinfo{year}{2019}).

\bibitem{2018MNRAS.477.2431T}
\bibinfo{author}{{Tarango-Yong}, J.~A.} \& \bibinfo{author}{{Henney}, W.~J.}
\newblock \bibinfo{title}{{True versus apparent shapes of bow shocks}}.
\newblock \emph{\bibinfo{journal}{\mnras}} \textbf{\bibinfo{volume}{477}},
  \bibinfo{pages}{2431--2454} (\bibinfo{year}{2018}).

\bibitem{2013MNRAS.431..279S}
\bibinfo{author}{{Sabin}, L.} \emph{et~al.}
\newblock \bibinfo{title}{{New Galactic supernova remnants discovered with
  IPHAS}}.
\newblock \emph{\bibinfo{journal}{\mnras}} \textbf{\bibinfo{volume}{431}},
  \bibinfo{pages}{279--291} (\bibinfo{year}{2013}).

\bibitem{2015ApJS..216...29B}
\bibinfo{author}{{Bovy}, J.}
\newblock \bibinfo{title}{{galpy: A python Library for Galactic Dynamics}}.
\newblock \emph{\bibinfo{journal}{\apjs}} \textbf{\bibinfo{volume}{216}},
  \bibinfo{pages}{29} (\bibinfo{year}{2015}).

\bibitem{2005SSRv..120..165B}
\bibinfo{author}{{Burrows}, D.~N.} \emph{et~al.}
\newblock \bibinfo{title}{{The Swift X-Ray Telescope}}.
\newblock \emph{\bibinfo{journal}{\ssr}} \textbf{\bibinfo{volume}{120}},
  \bibinfo{pages}{165--195} (\bibinfo{year}{2005}).

\bibitem{2020ApJS..247...54E}
\bibinfo{author}{{Evans}, P.~A.} \emph{et~al.}
\newblock \bibinfo{title}{{2SXPS: An Improved and Expanded Swift X-Ray
  Telescope Point-source Catalog}}.
\newblock \emph{\bibinfo{journal}{\apjs}} \textbf{\bibinfo{volume}{247}},
  \bibinfo{pages}{54} (\bibinfo{year}{2020}).

\bibitem{arnaud96}
\bibinfo{author}{{Arnaud}, K.~A.}
\newblock \bibinfo{editor}{{Jacoby}, G.~H.} \& \bibinfo{editor}{{Barnes}, J.}
  (eds) \emph{\bibinfo{title}{{XSPEC: The First Ten Years}}}.
\newblock (eds \bibinfo{editor}{{Jacoby}, G.~H.} \& \bibinfo{editor}{{Barnes},
  J.}) \emph{\bibinfo{booktitle}{Astronomical Data Analysis Software and
  Systems V}}, Vol. \bibinfo{volume}{101} of
  \emph{\bibinfo{series}{Astronomical Society of the Pacific Conference
  Series}}, \bibinfo{pages}{17} (\bibinfo{year}{1996}).

\bibitem{wilms00}
\bibinfo{author}{{Wilms}, J.}, \bibinfo{author}{{Allen}, A.} \&
  \bibinfo{author}{{McCray}, R.}
\newblock \bibinfo{title}{{On the Absorption of X-Rays in the Interstellar
  Medium}}.
\newblock \emph{\bibinfo{journal}{\apj}} \textbf{\bibinfo{volume}{542}},
  \bibinfo{pages}{914--924} (\bibinfo{year}{2000}).

\bibitem{HI4PI16}
\bibinfo{author}{{HI4PI Collaboration}} \emph{et~al.}
\newblock \bibinfo{title}{{HI4PI: A full-sky H I survey based on EBHIS and
  GASS}}.
\newblock \emph{\bibinfo{journal}{\aap}} \textbf{\bibinfo{volume}{594}},
  \bibinfo{pages}{A116} (\bibinfo{year}{2016}).

\bibitem{Struder01}
\bibinfo{author}{{Str{\"u}der}, L.} \emph{et~al.}
\newblock \bibinfo{title}{{The European Photon Imaging Camera on XMM-Newton:
  The pn-CCD camera}}.
\newblock \emph{\bibinfo{journal}{\aap}} \textbf{\bibinfo{volume}{365}},
  \bibinfo{pages}{L18} (\bibinfo{year}{2001}).

\bibitem{Turner01}
\bibinfo{author}{{Turner}, M.~J.~L.} \emph{et~al.}
\newblock \bibinfo{title}{{The European Photon Imaging Camera on XMM-Newton:
  The MOS cameras }}.
\newblock \emph{\bibinfo{journal}{\aap}} \textbf{\bibinfo{volume}{365}},
  \bibinfo{pages}{L27--L35} (\bibinfo{year}{2001}).

\bibitem{Mason01}
\bibinfo{author}{{Mason}, K.~O.} \emph{et~al.}
\newblock \bibinfo{title}{{The XMM-Newton optical/UV monitor telescope}}.
\newblock \emph{\bibinfo{journal}{\aap}} \textbf{\bibinfo{volume}{365}},
  \bibinfo{pages}{L36} (\bibinfo{year}{2001}).

\bibitem{gaiadr3}
\bibinfo{author}{{Gaia Collaboration}} \emph{et~al.}
\newblock \bibinfo{title}{{Gaia Data Release 3. Summary of the content and
  survey properties}}.
\newblock \emph{\bibinfo{journal}{\aap}} \textbf{\bibinfo{volume}{674}},
  \bibinfo{pages}{A1} (\bibinfo{year}{2023}).

\bibitem{Tiengo2010}
\bibinfo{author}{{Tiengo}, A.} \emph{et~al.}
\newblock \bibinfo{title}{{The Dust-scattering X-ray Rings of the Anomalous
  X-ray Pulsar 1E 1547.0-5408}}.
\newblock \emph{\bibinfo{journal}{\apj}} \textbf{\bibinfo{volume}{710}},
  \bibinfo{pages}{227--235} (\bibinfo{year}{2010}).

\bibitem{jonas16}
\bibinfo{author}{{Jonas}, J.} \& \bibinfo{author}{{MeerKAT Team}}
\newblock \emph{\bibinfo{title}{{The MeerKAT Radio Telescope}}},
  \bibinfo{pages}{1} (\bibinfo{year}{2016}).

\bibitem{oxcat}
\bibinfo{author}{Heywood, I.}
\newblock \bibinfo{title}{{oxkat}: Semi-automated imaging of meerkat
  observations}.
\newblock \bibinfo{howpublished}{Astrophysics Source Code Library, record
  ascl:2009.003} (\bibinfo{year}{2020}).

\bibitem{CASAteam2022}
\bibinfo{author}{{CASA Team}} \emph{et~al.}
\newblock \bibinfo{title}{{CASA, the Common Astronomy Software Applications for
  Radio Astronomy}}.
\newblock \emph{\bibinfo{journal}{\pasp}} \textbf{\bibinfo{volume}{134}},
  \bibinfo{pages}{114501} (\bibinfo{year}{2022}).

\bibitem{tricolour}
\bibinfo{author}{{Hugo}, B.~V.}, \bibinfo{author}{{Perkins}, S.},
  \bibinfo{author}{{Merry}, B.}, \bibinfo{author}{{Mauch}, T.} \&
  \bibinfo{author}{{Smirnov}, O.~M.}
\newblock \bibinfo{editor}{{Ruiz}, J.~E.}, \bibinfo{editor}{{Pierfedereci}, F.}
  \& \bibinfo{editor}{{Teuben}, P.} (eds) \emph{\bibinfo{title}{{Tricolour: An
  Optimized SumThreshold Flagger for MeerKAT}}}.
\newblock (eds \bibinfo{editor}{{Ruiz}, J.~E.},
  \bibinfo{editor}{{Pierfedereci}, F.} \& \bibinfo{editor}{{Teuben}, P.})
  \emph{\bibinfo{booktitle}{Astronomical Data Analysis Software and Systems
  XXX}}, Vol. \bibinfo{volume}{532} of \emph{\bibinfo{series}{Astronomical
  Society of the Pacific Conference Series}}, \bibinfo{pages}{541}
  (\bibinfo{year}{2022}).
\newblock
  \bibinfo{eprint}{{\href{https://arxiv.org/abs/2206.09179}{{arXiv:2206.09179}}}}.

\bibitem{wsclean}
\bibinfo{author}{Offringa, A.~R.}, \bibinfo{author}{McKinley, B.},
  \bibinfo{author}{Hurley-Walker} \emph{et~al.}
\newblock \bibinfo{title}{{WSClean: an implementation of a fast, generic
  wide-field imager for radio astronomy}}.
\newblock \emph{\bibinfo{journal}{MNRAS}} \textbf{\bibinfo{volume}{444}},
  \bibinfo{pages}{606--619} (\bibinfo{year}{2014}).

\bibitem{cubical}
\bibinfo{author}{{Kenyon}, J.~S.}, \bibinfo{author}{{Smirnov}, O.~M.},
  \bibinfo{author}{{Grobler}, T.~L.} \& \bibinfo{author}{{Perkins}, S.~J.}
\newblock \bibinfo{title}{{CUBICAL - fast radio interferometric calibration
  suite exploiting complex optimization}}.
\newblock \emph{\bibinfo{journal}{\mnras}} \textbf{\bibinfo{volume}{478}},
  \bibinfo{pages}{2399--2415} (\bibinfo{year}{2018}).

\bibitem{2012ascl.soft08017R}
\bibinfo{author}{Robitaille, T.} \& \bibinfo{author}{Bressert, E.}
\newblock \bibinfo{title}{{APLpy}: Astronomical plotting library in python}.
\newblock \bibinfo{howpublished}{Astrophysics Source Code Library, record
  ascl:1208.017} (\bibinfo{year}{2012}).

\bibitem{MUSE_cube}
\bibinfo{author}{Ilkiewicz, K.} \emph{et~al.}
\newblock \bibinfo{title}{Muse observation of 1rxs j052832.5+283824 bow shock}
  (\bibinfo{year}{2025}).
\newblock \urlprefix\url{https://doi.org/10.5281/zenodo.17238003}.

\end{thebibliography}
\end{document}